\documentclass[fleqn,usenatbib]{mnras}

\usepackage{newtxtext,newtxmath}

\usepackage[T1]{fontenc}

\DeclareRobustCommand{\VAN}[3]{#2}
\let\VANthebibliography\thebibliography
\def\thebibliography{\DeclareRobustCommand{\VAN}[3]{##3}\VANthebibliography}

\usepackage{graphicx}	
\usepackage{amsmath}	
\usepackage{rotating}
\usepackage{caption}
\usepackage[dvipsnames]{xcolor}

\newcommand{\msun}{$M_{\odot}$}

\newcommand{\sigmagas}{$\Sigma_\text{gas}$}
\newcommand{\sigmadust}{$\Sigma_\text{dust}$}

\newcommand{\gasdust}{$\Delta_\text{g/d}$}
\newcommand{\rsub}{$R_\text{sub}$}
\newcommand{\rgap}{$R_\text{gap}$}
\newcommand{\rcav}{$R_\text{cav}$}

\newcommand{\citeblue}[1]{{\color{blue}{#1}\color{black}}}

\title[Volatiles in HD 169142]{Volatile composition of the HD 169142 disk and its embedded planet}

\author[Luke Keyte]{
Luke Keyte,$^{1}$\thanks{E-mail: luke.keyte.18@ucl.ac.uk}
Mihkel Kama,$^{1,2}$
Alice S. Booth,$^{3}$
Charles J. Law,$^{4}$
Margot Leemker$^{5}$
\\
$^{1}$Department of Physics and Astronomy, University College London, Gower Street, WC1E 6BT London, United Kingdom\\
$^{2}$Tartu Observatory, University of Tartu, Observatooriumi 1, 61602 Toravere, Tartumaa, Estonia\\
$^{3}$Center for Astrophysics | Harvard \& Smithsonian, 60 Garden St., Cambridge, MA 02138, USA\\
$^{4}$Department of Astronomy, University of Virginia, Charlottesville, VA 22904, USA\\
$^{5}$Leiden Observatory, Leiden University, 2300 RA Leiden, the Netherlands\\
}

\date{Accepted 2024 October 01. Received 2024 September 04; in original form 2024 May 22}

\pubyear{2024}

\begin{document}
\label{firstpage}
\pagerange{\pageref{firstpage}--\pageref{lastpage}}
\maketitle

\begin{abstract}

The composition of a planet's atmosphere is intricately linked to the chemical makeup of the protoplanetary disk in which it formed. Determining the elemental abundances from key volatiles within disks is therefore essential for establishing connections between the composition of disks and planets. The disk around the Herbig Ae star HD\;169142 is a compelling target for such a study due to its molecule-rich nature and the presence of a newly-forming planet between two prominent dust rings. In this work, we probe the chemistry of the HD\;169142 disk at small spatial scales, drawing links between the composition of the disk and the planet-accreted gas. Using thermochemical models and archival data, we constrain the elemental abundances of volatile carbon, oxygen, and sulfur. Carbon and oxygen are only moderately depleted from the gas phase relative to their interstellar abundances, with the inner \char126 60 au appearing enriched in volatile oxygen. The C/O ratio is approximately solar within the inner disk (\char126 0.5) and rises above this in the outer disk (>0.5), as expected across the H$_2$O snowline. The gas-phase sulfur abundance is depleted by a factor of \char126 1000, consistent with a number of other protoplanetary disks. Interestingly, the observed SiS emission near the HD\;169142 b protoplanet vastly exceeds chemical model predictions, supporting previous hypotheses suggesting its origin in shocked gas or a localised outflow. We contextualise our findings in terms of the potential atmospheric composition of the embedded planet, and highlight the utility of sulfur-bearing molecules as probes of protoplanetary disk chemistry.


\end{abstract}



\section{Introduction}

Planets are born within the disks of gas, dust, and ice that surround young stars. These ingredients serve as the fundamental building blocks from which newly-forming planets assemble. Insights into the chemical composition of such disks are vital to unlocking the mysteries that surround planetary origins, in our own Solar System and beyond. Over the past decade, sub-millimetre observations with Atacama Large Millimeter/submillmeter Array (ALMA) have revealed protoplanetary disks to be complex environments, with physical and chemical substructures that vary at small spatial scales \citep{Andrews_2018, huang_2018, maps_1_oberg2021, maps_3_law2021}. Rings and gaps are commonly observed in the distribution of gas and dust, and are often linked to dynamical interactions between the disk and embedded planets \citep[e.g.][]{dong_2015, Andrews_2018, long_2018}. However, a wide range of other mechanisms have also been proposed to explain such features, which include dead-zones induced by the magneto-rotational instability \citep{flock_2015}, pebble growth at molecular snowlines \citep{zhang_2015, booth_2017}, and photoevaporation \citep{vellejo_2021}. Drawing links between individual substructures and newly-forming planets is therefore not straightforward.

The abundance and distribution of volatile elements within these disks offer valuable insights into the physical and chemical processes that govern planetary formation, while also shedding light on the underlying mechanisms responsible for creating observed substructures. Multi-wavelength observations have shown that protoplanetary disks are chemically-rich environments, with a wide range of simple and complex molecular species having been detected thus far \citep[e.g.][]{pontoppidan_2010, meeus2012, mcguire_2022, oberg_2023}. While broad compositional trends have been identified across the general protoplanetary disk population, the specific makeup of individual sources can often vary significantly. Various techniques have been employed to characterise disk chemistry, the most ubiquitous of which is to place constraints on the carbon-to-oxygen ratio (C/O). Both carbon and oxygen are among the most abundant volatile elements in disks and planetary atmospheres, and  the C/O ratio is often treated as a probe of formation location, establishing a link between the atmospheric composition of a planet and the region of the disk in which accretion occurred \citep[e.g.][]{oberg2011}. The gas-phase C/O ratio in disks is often estimated by combining spatially resolved observations of C$_2$H emission with sophisticated chemical models to derive carbon and oxygen abundances. Numerous studies which adopt this method have found C/O to be enhanced in the outer regions of disks, often coupled with the depletion of volatile carbon and oxygen \citep[e.g.][]{miotello_2019, maps_7_bosman2021}. A simple understanding of these findings invokes snowlines of key molecules, whereby freezeout leads to a radially increasing gas-phase C/O profile. However, additional mechanisms such as radial drift, dust trapping, turbulence, and disk/planetesimal evolution substantially complicate this picture \citep[e.g.][]{Bergin_2016,booth_2017, Krijt2018, booth_ilee_2019, Van_Clepper_2022}. Connecting atmospheric C/O ratios to precise planet formation locations therefore remains challenging.

Beyond C/O, other elemental ratios such as C/N, N/O, and S/N, have proven useful in probing disk chemistry \citep{turrini_2021}. Sulfur is of particular interest, with sulfur-bearing molecules now being routinely detected in disks \citep{dutrey_2011, guilloteau_2013, phuong_2018, legal_2019, Booth_2021, law_2023, booth_2024a, booth_2024b}, and observable in exoplanet atmospheres with the James Webb Space Telescope (JWST) \citep{rustamkulov_2023, tsai_2023}. Although chemical models constrained by spectroscopic observations suggest that sulfur is heavily depleted from the gas phase in planet-forming environments \citep{fuente_2010, legal_2019, maps_12_legal2021, keyte_2024}, emission from sulfur-bearing species has still been harnessed as a useful probe of protoplanetary disk gas. The CS/SO ratio, for example, can be employed as a proxy for the C/O ratio, since CS/SO varies by many orders of magnitude for small variations in C/O. Studies that utilise the CS/SO ratio to probe disk gas have revealed extreme diversity amongst disks, with both oxygen-dominated and carbon-dominated chemistry having been observed \citep[e.g.][]{Semenov_2018, Booth_2021, keyte_2023, booth_2023b}. Observations of individual species such as SO and SiS have also found use as tracers of dynamical features such as disk winds \citep{lee_2021}, probes of shocked gas in the vicinity of nascent planets \citep{booth_2023b}, and as evidence for dynamical heating \citeblue{(Zhang et al. 2024)}. Sulfur-bearing molecules therefore have wide utility as probes of the complex dynamics and chemistry within planet-forming environments.

One such system in which a range of sulfur-bearing species have been detected is the HD\;169142 protoplanetary disk. In this paper, we present thermochemical models of this system, which are constrained using a range of high-resolution archival data. Our aim is to determine the gas-phase abundance of elemental carbon, oxygen, and sulfur at the small spatial scales relevant to planet formation. We place particular focus on modelling individual sulfur-bearing molecules, and investigate links between the molecular emission and a proposed embedded protoplanet HD\;169142 b. The archival observations are described in Section\,\ref{sec: observations} and the details of the model are outlined in Section\,\ref{sec: modelling}. We present our results in Section\,\ref{sec: results} and discuss their implications in Section\,\ref{sec: discussion}. We summarize our conclusions in Section\,\ref{sec: conclusions}.

\section{Observations}
\label{sec: observations}

\subsection{The HD\;169142 disk}

HD\;169142 is a well-studied nearby Herbig Ae star located at a distance of $d=115$ pc, with an estimated mass of $M_* = 1.65 M_\odot$ \citep{blondel_djie_2006, bailer_jones_2021, gaia_collaboration_2021}. The stellar luminosity is $L_* = 10 L_\odot$ \citep{fedele_2017} and effective temperature $T_*=8400$ K \citep{dunkin_1997}. The star is host to an almost face-on protoplanetary disk, with inclination $i=(13\pm1)^\circ$ and position angle PA $=(5\pm5)^\circ$ \citep{raman_2006, panic_2008}.

\begin{figure}
\centering
\includegraphics[clip=,width=1.0\linewidth]{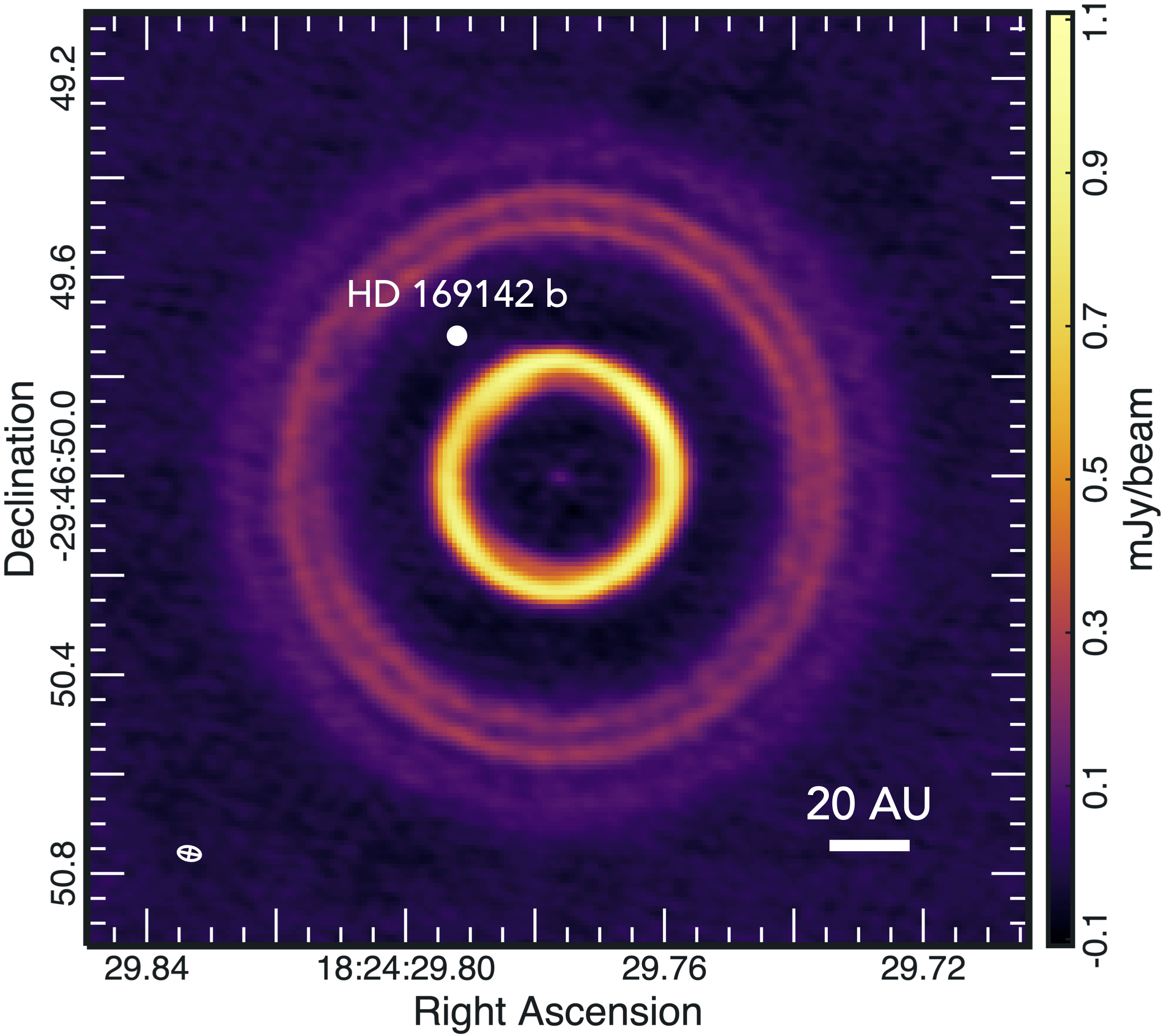}
\caption{ALMA 1.3mm continuum observation of the HD 169142 system \citep{perez_2019}, including the inferred location of the HD 169142 b protoplanet \citep{hammond_2023}. The synthesized beam is indicated by the white ellipse in the lower left.}
\label{fig_continuum_emission}
\end{figure}

Observations of dust at infrared and sub-millimeter wavelengths have revealed a number of well-defined rings and gaps, which have been linked to ongoing planetary formation in the disk \citep[e.g.][]{quanz_2013, fedele_2017, macias_2019, pohl_2017, perez_2019, bertrang_2018}. A wide range of planetary masses and locations have been proposed in the literature, with particular attention given to a candidate located in the outer disk between two dust rings at $r\sim 37$ au. Hydrodynamic models lend support to the planetary origin of this gap \citep{toci_2020}, while deviations from the Keplerian motion of CO gas provide additional evidence \citep{garg_2022}.  Multi-epoch imaging at near-infrared (NIR) wavelengths revealed a hotspot of emission in the proposed planetary location, which exhibits clockwise motion consistent with a Keplerian orbit \citep{gratton_2019, hammond_2023}. We highlight the ringed structure of the disk in Figure \ref{fig_continuum_emission}, which shows an ALMA observation of 0.89 mm continuum emission \citep{perez_2019}, alongside the proposed location of the protoplanet HD\;169142 b.

The disk is also rich in atomic and molecular line data. A wide range of species have been detected, including CO (+isotopologues), deuterated and sulfur-bearing molecules, formaldehyde, and methanol \citep[e.g.][]{panic_2008, meeus2012, carney_2018, yu_2021, leemker_2022, booth_2023b}. Recent findings include the first detection of SiS in a protoplanetary disk, which alongside SO, has been proposed to trace shocked gas in the proximity of the aforementioned protoplanet \citep{law_2023}.

\subsection{Archival data}

In this study, we make use of a wealth of archival data. This includes high-resolution continuum observations from ALMA at 0.45\,mm \citep{leemker_2022}, 0.89\,mm \citep{law_2023}, and 1.3\,mm \citep{garg_2022}, and spectral line observations from ALMA covering a range of chemical species: CO(+isotopologues, [CI], CS(+isotopologues), SO, H$_2$CS, and SiS \citep{leemker_2022, garg_2022, law_2023, booth_2023a}. We also incorporate disk-integrated fluxes and upper-limits of HD 56$\mu$m/112$\mu$m and [OI] 63$\mu$m/146$\mu$m from Herschel/PACS \citep{meeus2012, kama_2020}. The data is outlined in detail in Table \ref{table:observational_data}.

\section{Modelling}
\label{sec: modelling}

We ran source-specific models using the 2D physical-chemical code \textsc{dali} \citep{Bruderer2012, Bruderer2013}. The code first requires a parameterised gas and dust density distribution and an input stellar spectrum, then uses Monte Carlo radiative transfer to determine the UV radiation field and dust temperature. This provides an initial guess for the the gas temperature, which begins an iterative process in which the chemistry is solved time-dependently. We let the chemistry evolve to 5 Myr, consistent with recent age estimates for the system \citep{Vioque_2018}. Finally, the raytracing module is used to obtain spectral image cubes, line profiles, and disk-integrated line fluxes.

In this work, we use the disk parameters from \citet{fedele_2017} and \citet{carney_2018} as our starting point for refining the model. Incorporating a wider range of observational constraints, we determine the total elemental carbon and oxygen abundances, then extend the model to include a range of sulfur-bearing molecules. Details of the model are described below.

\subsection{Disk structure}

The disk structure is fully parameterised, with a surface density that follows the standard form of a power law with an exponential taper:
\begin{equation}
    \Sigma_\text{gas} = \Sigma_\text{c} \cdot \bigg(\frac{r}{R_c} \bigg)^{-\gamma} \exp \bigg[- \bigg(\frac{r}{R_c} \bigg)^{2-\gamma} \bigg]
\end{equation}
where $r$ is the radius, $\gamma$ is the surface density exponent, $\Sigma_\text{c}$ is some critical surface density, and $R_\text{c}$ is some critical radius, such that the surface density at $R_\text{c}$ is $\Sigma_\text{c}/e$. The scale height is then given by:
\begin{equation}
    h(r) = h_c\bigg(\frac{r}{R_c}\bigg)^\psi
\end{equation}
where $h_\text{c}$ is the scale height at $R_\text{c}$, and the power law index of the scale height, $\psi$, describes the flaring of the disk.

\sigmagas \ and \sigmadust \ extend from the dust sublimation radius (\rsub, defined as the inner edge of the model domain) to the edge of the disk ($R_\text{out}$). The gas and dust surface densities can be varied independently within a number of predefined radial regions, using the various $\delta_\text{gas}$ and $\delta_\text{dust}$ parameters (see Table \ref{table:modelparameters} for a full description).

The gas-to-dust ratio is denoted \gasdust. Dust settling is implemented by considering two different populations of grains: small grains (0.005$\mu$m -1$\mu$m) and large grains (0.005$\mu$m -1mm). The size distribution of both populations is proportional to $a^{-3.5}$ (where $a$ is the grain size), consistent with the interstellar grain size distribution \citep{mathis_1977}. The vertical density structure of the dust is such that large grains are settled towards the midplane, prescribed by the settling parameter $\chi$:

\begin{equation}
    \rho_\text{dust, large} = \frac{f \Sigma_\text{dust}}{\sqrt{2\pi}r \chi h} \exp \bigg[ -\frac{1}{2} \bigg( \frac{\pi/2 - \theta}{ \chi h} \bigg) ^{2} \bigg]
\end{equation}

\begin{equation}
    \rho_\text{dust, small} = \frac{(1-f)\Sigma_\text{dust}}{\sqrt{2\pi}rh} \exp \bigg[ -\frac{1}{2} \bigg( \frac{\pi/2 - \theta}{h} \bigg) ^{2} \bigg]
\end{equation}
where $f$ is the mass fraction of large grains and $\theta$ is the opening angle from the midplane as viewed from the central star.

\subsection{Stellar parameters}
The stellar spectrum is modelled as a blackbody with temperature $T=8400$ K \citep{dunkin_1997} and luminosity $L_*=10L_\odot$ \citep{fedele_2017}. The X-ray spectrum was characterised as a thermal spectrum with a temperature of $7 \times 10^7$ K over the 1 to 100 keV energy range, with an X-ray luminosity of $L_X = 7.94 \times 10^{28}$ erg s$^{-1}$, based on observations of similar Herbig Ae/Be systems \citep{stelzer_2006}.

\subsection{Chemical networks}

To constrain the overall carbon and oxygen abundances, we employ the CO isotopologue network from \citet{miotello_2016}, which is based on a subset of the UMIST 06 \citep{woodall2007} network. This network includes CO isotopologues, allowing us to take into account the effects of isotope-selective photodissociation, where the isotope ratios are taken to be [$^{12}$C]/[$^{13}$C] = 77 and [$^{16}$O]/[$^{18}$O] = 560 \citep{wilson_rood_1994}. In total, the CO isotopologue network consists of 185 species (including neutral and charged PAHs) and 5755 individual reactions. The code includes H$_2$ formation on dust, freeze-out, thermal desorption, hydrogenation, gas-phase reactions, photodissociation and photoionisation, X-ray induced processes, cosmic-ray induced reactions, PAH charge exchange/hydrogenation and reactions with vibrationally excited H$_2$. Non-thermal desorption is only included for a small number of species (CO, CO$_2$, H$_2$O, CH$_4$, NH$_3$). For grain surface chemistry, only hydrogenation of simple species is considered (C, CH, CH$_2$, CH$_3$, N, NH, NH$_2$, O, OH). The details of these processes are described more fully in \citet{Bruderer2012}. 

Since the CO isotopologue network is missing important sulfur reactions, we use the network from \citet{keyte_2023} to model the sulfur-bearing species. This network augments the chemical network from \citet{Bruderer2013} with additional sulfur-bearing molecules and reactions. These include gas-phase reactions from UMIST 06, freezeout and thermal desorption, photoionisation and photodissociation, and hydrogenation of S to HS and HS to H$_2$S on grain surfaces. The network was previously tested against a model incorporating the full set of gas-phase reactions from the RATE 12 dataset (467 species, 5950 reactions; \citealt{mcelroy_umist12}) using the \textsc{starchem} modelling code (\citealp{rawlings_2022}, Rawlings, Keto \& Caselli, submitted), finding good agreement (see \citealt{keyte_2024} for full details). For this study, we extended the network to include two additional sulfur-bearing species, S$_2$ and S$^-$, which were found to play a role in the production of OCS and SO in our previous study of sulfur chemistry in the HD\;100546 protoplanetary disk \citep{keyte_2024}. We also implemented non-thermal desorption for all sulfur-bearing species, where the photodesorption yield is set to $Y=1.2 \times 10^{-3}$ for H$_2$S \citep{fuente_2017}, and $Y=10^{-3}$ for all other species (for which the yields have not been experimentally determined). This value is consistent with that used in previous studies of photodesorption where the yield is not known \citep[e.g.][]{visser_2011}. In total, the network used for sulfur chemistry consists of 135 individual species and 1775 individual reactions. To maintain consistency, we adopt the initial abundances and binding energies from the CO isotopologue network. A similar approach has previously been used to model sulfur chemistry in the disk around DR Tau \citep{huang_2024}.

\section{Results}
\label{sec: results}

\subsection{Baseline model}

\begin{figure*}
\centering
\includegraphics[clip=,width=1.0\linewidth]{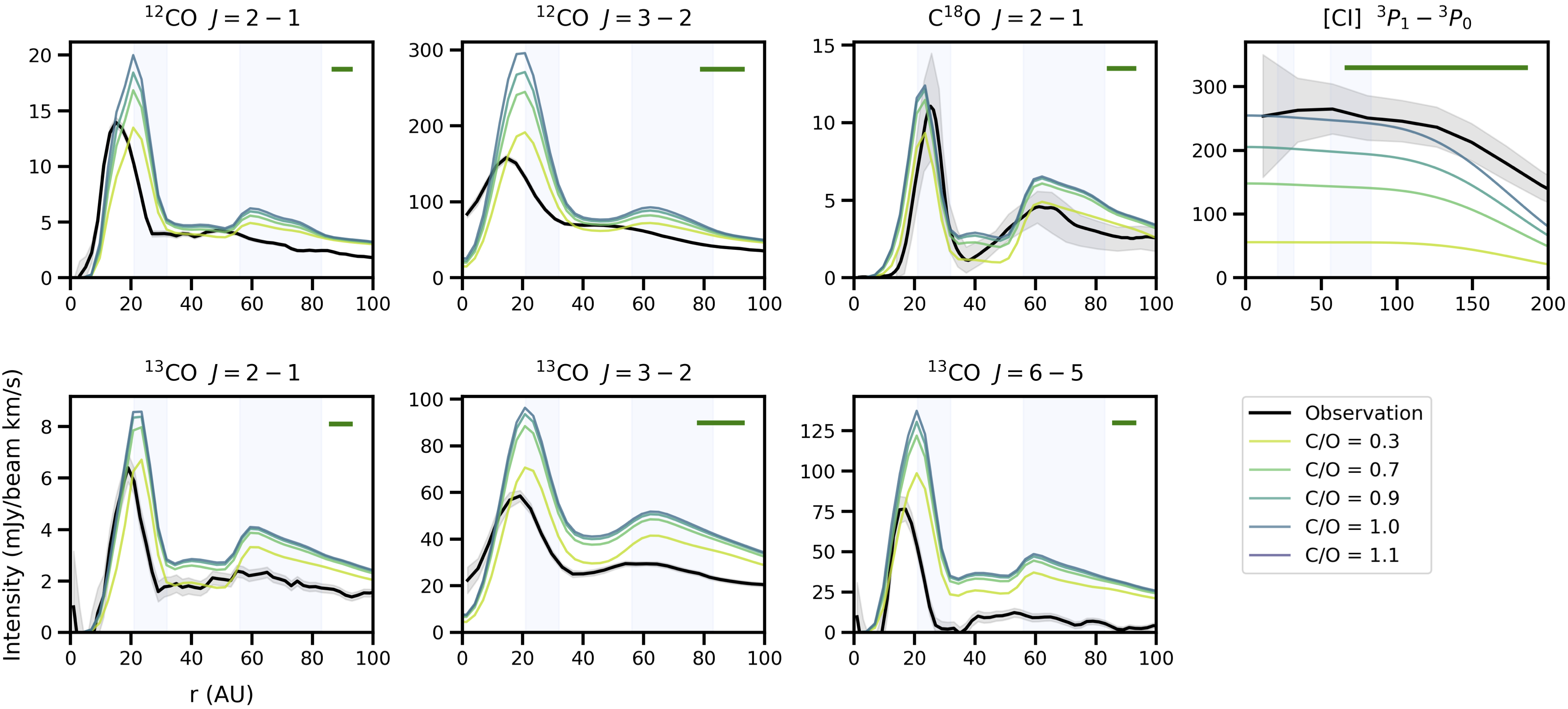}
\caption{Radial intensity profiles for the CO (+isotopologues) and [CI] line listed in Table \ref{table:observational_data}. Observations are in black, with uncertainties shaded in grey. Models with varying C/O ratios are shown as coloured lines. Each model uses a fixed initial oxygen abundance of $2 \times 10^{-4}$, with the C/O ratio controlled by variations in C/H. The models are otherwise identical. All models are convolved with a beam to match the observations, where the beam size is indicated by the green bar in the top right of each panel. Shaded blue regions denote dust rings locations.}
\label{fig_radial_coratio}
\end{figure*}

\begin{figure*}
\centering
\includegraphics[clip=,width=1.0\linewidth]{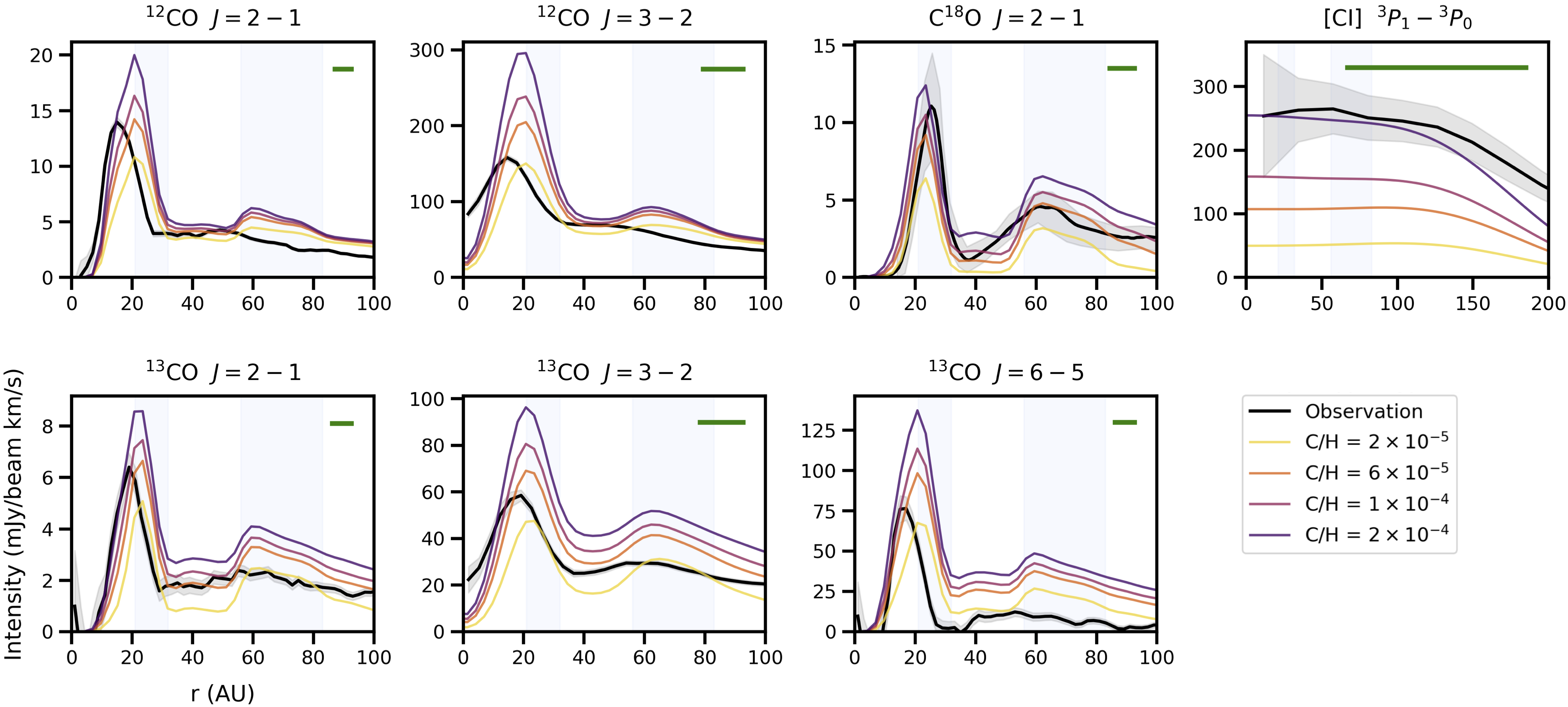}
\caption{Radial intensity profiles for the CO (+isotopologues) and [CI] line listed in Table \ref{table:observational_data}. Observations are in black, with uncertainties shaded in grey. Models with varying C/H shown as coloured lines. Each model uses a fixed C/O ratio of 1.0, where O/H is varied in step with C/H. The models are otherwise identical. All models are convolved with a beam to match the observations, where the beam size is indicated by the green bar in the top right of each panel. Shaded blue regions denote dust rings locations.}
\label{fig_radial_ch_oh}
\end{figure*}

Our fitting procedure follows the approach of \citet{Kama2016b} and \citet{keyte_2023}. We begin by constraining the dust distribution by jointly fitting the spectral energy distribution and high-resolution continuum observations. Using a grid of $\sim2500$ models, we vary the parameters $\psi$, $h_\text{c}$, $\Delta_\text{g/d}$, $\chi$, $f$, adopting the values from \citet{fedele_2017} as a starting point. We also include variations in the parameters that define the radial location of the dust rings and the level of dust depletion within the gaps. At this stage, \sigmagas \;is kept fixed at an arbitrary value such that changes to \gasdust\ are equivalent only to changes in the dust mass. 

The results for our best-fitting model are illustrated in Figure \ref{fig_continuum_sed}. The best-fitting parameter values are consistent with \citet{fedele_2017}, with the only difference being that our model incorporates a slightly narrower inner dust ring, extending from 21 to 32 au, rather than 20 to 35 au, which is constrained using higher resolution data. The model provides an excellent match to the spectral energy distribution and 0.89mm continuum radial intensity profile, and matches the 0.45mm and 1.3mm continuum fluxes to within a factor of $\sim 1.3$. In an effort to improve the fit, we explored the impact of varying additional model parameters such as the minimum/maximum grain sizes, and the power law index of the grain size distribution. However, we were unable to simultaneously decrease the 0.45mm continuum flux while increasing the 1.3mm flux, which may be suggestive of a more complex grain size distribution or different dust opacities to those used in our model. The total dust mass is $\sim 10^{-4}\;M_\odot$, which is again consistent with previous studies.

Next, we use the upper limits of the HD 56$\mu$m and 112$\mu$m disk-integrated fluxes to constrain the maximum gas mass. A second grid of models is run in which \sigmagas and \gasdust\ are varied simultaneously, allowing the gas mass to vary while maintaining the best-fit dust mass. Our model grid covers gas-to-dust ratios ranging from 10 to 500, but the HD upper limits are not tight enough to provide any useful constraint. From this point on, we therefore adopt the interstellar value of $\Delta_\text{g/d} = 100$.

\begin{figure}
\centering
\includegraphics[clip=,width=0.95\linewidth]{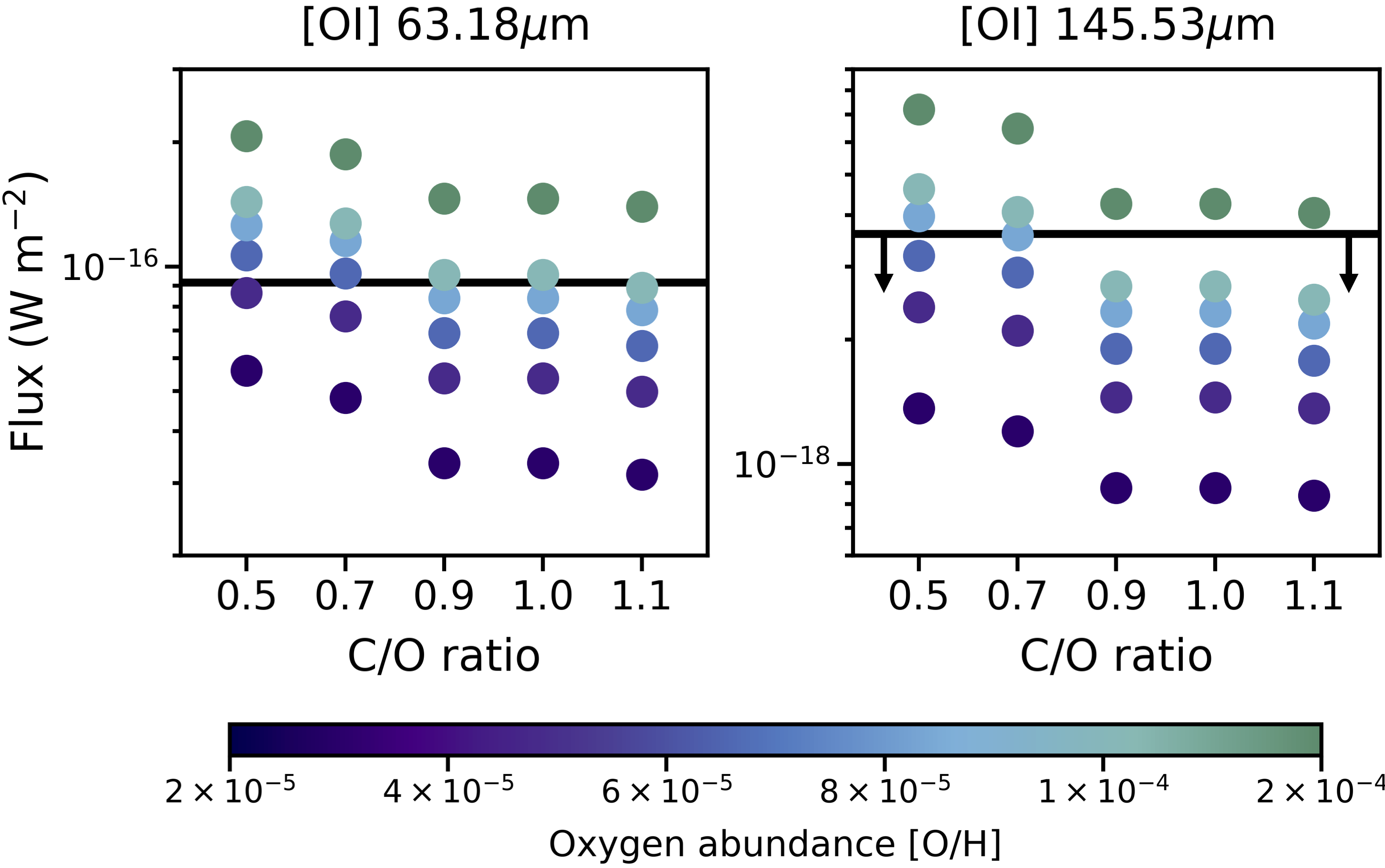}
\caption{[OI] 63.18$\mu$m and 145.53$\mu$m disk-integrated line fluxes. Observations are denoted by the black lines, where [OI] 63.18$\mu$m is a detection and [OI] 145.53$\mu$m is a 3$\sigma$ upper limit. Modelled fluxes cover a range of C/O ratios (\emph{x}-axis) and initial oxygen abundances ranging from O/H$=2\times 10^{-5}$ to $2 \times 10^{-4}$ (colors).}
\label{fig_oxygen}
\end{figure}

The total elemental carbon and oxygen abundances, C/H and O/H, are constrained by modelling the CO(+isotopologues) and [CI] radial intensity profiles, as well as the [OI] disk-integrated line fluxes. We focus initially on fitting the C$^{18}$O $J=2-1$ radial intensity profile, since this line is expected to be the most optically thin, and is therefore the most robust tracer of the overall gas surface density. We run models covering a range of C/O ratios (C/O = 0.3 to 1.1), and total elemental carbon and oxygen abundances (C/H and O/H between $2 \times 10^{-5}$ and $2 \times 10^{-4}$).

Figure \ref{fig_radial_coratio} illustrates a subset of models for which the C/O ratio is varied between 0.3 and 1.1, while the total elemental oxygen abundance is kept fixed at O/H $=2 \times 10^{-4}$. The radial intensity profile emission morphology is well-matched by all transitions, with the C$^{18}$O $J=2-1$ profile also providing a close match to the absolute intensity for all values of C/O. However, the peak intensity of the remaining CO isotopologues is slightly overpredicted in nearly all cases, typically by a factor $\lesssim 1.5$. The [CI] $^3P_1-^3P_0$ radial intensity profile is best matched by a model with a relatively high C/O ratio of 1.0, but increasing C/O above unity leads to significant overprediction of the [CI] $^3P_1-^3P_0$ profile (factor $>2$).

In Figure \ref{fig_radial_ch_oh} we show a second subset of models, this time varying the total elemental carbon and oxygen abundances, while keeping the C/O ratio fixed at C/O=1.0. A relatively high carbon abundance (C/H $\gtrsim 1 \times 10^{-4}$) is required to match the peak intensity of the C$^{18}$O $J=2-1$ emission, and the [CI] $^3P_1-^3P_0$ radial intensity profile is only matched when using the maximum value C/H$=2 \times 10^{-4}$. However, adopting a relatively high total elemental carbon abundance once again leads to overprediction of the remaining CO isotopologues. In particular, the modelled $^{13}$CO intensity profiles are overpredicted in the outer disk by larger factors for higher-$J$ transitions. 

Further constraints come from the [OI] disk-integrated fluxes and upper limits, illustrated in Figure \ref{fig_oxygen},  The observations are shown alongside a range of models spanning C/O=0.5 to 1.1 and O/H = $2\times 10^{-5}$ to $2\times 10^{-4}$. When C/O $\gtrsim 0.9$, the observations are best fit by models with O/H\;$\sim 10^{-4}$, consistent with the [CI] $^3P_1-^3P_0$ data. When C/O $\lesssim 0.9$, lower oxygen abundances are required (O/H\;$\sim 5\times 10^{-5}$), which corresponds to similarly reduced carbon abundances. However, these models are inconsistent with the [CI] $^3P_1-^3P_0$ data.

Determining best-fitting values for C/H and O/H is challenging. The C$^{18}$O, [CI], and [OI] data point towards low levels of carbon/oxygen depletion (C/H and O/H $\sim 10^{-4}$  i.e. approximately the full interstellar abundance) and a C/O ratio close to, but not exceeding, unity. However, models incorporating these values typically overpredict emission from the $^{12}$CO and $^{13}$CO isotopologues, by a factor that increases for higher-$J$ transitions. However, these lines are expected to be optically thick, and as such their emission predominantly traces the gas temperature in the emitting region rather than variations in the elemental abundance(s). The gas/dust temperature/density structures from our model are presented in Figure \ref{fig_tempdens}, which closely match models from previous studies \citep{fedele_2017, carney_2018}, since the density structures and stellar parameters are extremely similar. Consistent overprediction of the $^{12}$CO and $^{13}$CO emission suggests that the modelled gas temperature in the disk atmosphere, from where these species mainly emit, is too high. We first investigated whether this may be related to the modelled stellar properties by exploring variations in the stellar luminosity, stellar temperature, and stellar X-ray spectrum. However, varying each of these parameters was found to significantly impact the SED fit for only minimal variations in the $^{12}$CO and $^{13}$CO emission. Similarly, small variations in the disk physical structure parameters produce only small variations in the $^{12}$CO/$^{13}$CO emission while strongly affecting the SED. Manually reducing the gas temperature in the disk atmosphere by 10 K was also found to strongly impact the SED, while only reducing the $^{12}$CO/$^{13}$CO emission by $\sim8$\%. We note, however, that if the true gas temperature in the CO emitting layer is indeed slightly lower than our model, this would require slightly higher C$^{18}$O and atomic carbon/oxygen abundances to reproduce the C$^{18}$O, [CI], and [OI] emission. Thus, regardless of the precise cause of the optically thick CO emission overprediction, our conclusion that C and O are undepleted in the gas still holds.

\begin{figure*}
\centering
\includegraphics[clip=,width=1.0\linewidth]{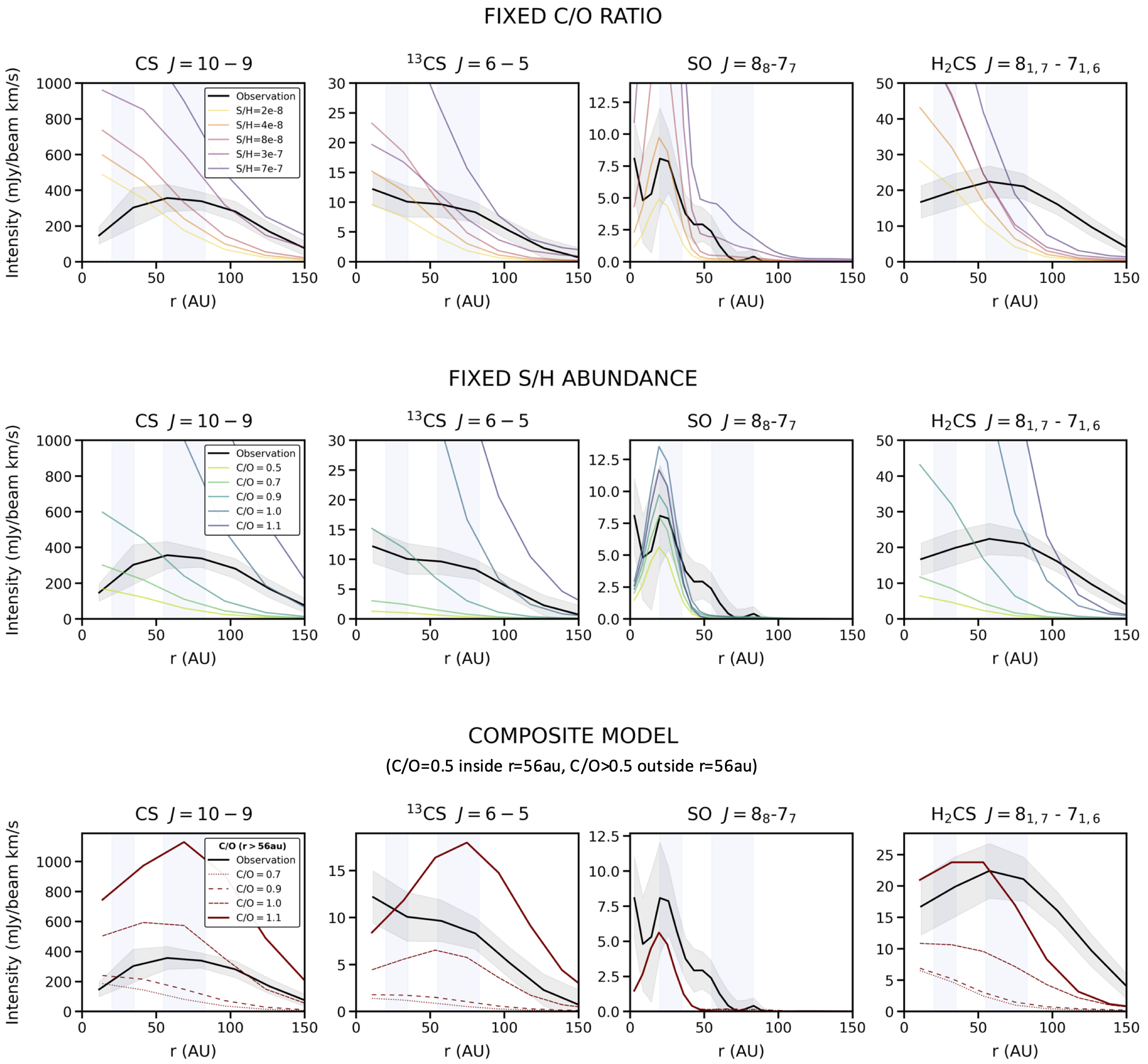}
\caption{Radial intensity profiles for the CS $J=10-9$, $^{13}$CS $J=6-5$, SO $J=8_8-7_7$, and H$_2$CS $J=8_{1,7}-7_{1,6}$ transitions (observations in black and models in color). Blue shaded regions denote the location of dust rings. \emph{Top row: }Models with varying initial elemental sulfur abundance S/H = $2 \times 10^{-8}$ to $10^{-7}$, where C/O is fixed at 0.9 and C/H = $10^{-4}$. \emph{Middle row: }Models with varying C/O ratio (0.3 to 0.9), where the total elemental sulfur abundance is fixed at S/H $=4\times 10^{-8}$. \emph{Bottom row: }Composite models, where C/O=0.5 inside of the outer dust ring ($r=56$ au), and C/O varies outside this region according to the legend. The total elemental sulfur abundance is again fixed at S/H $=4\times 10^{-8}$. No change is seen in the SO profile, since the C/O variations are only present beyond $r=56$ au where the SO emission is effectively zero.}
\label{fig_sulfur_grid}
\end{figure*}

Adopting C/O $=1.0$ and C/H $=1\times 10^{-4}$ as fiducial values, we use the C$^{18}$O data to constrain the level of gas depletion in different radial regions throughout the disk. Inside the inner dust ring ($r \sim 21-32$ au), we are only able to reproduce the C$^{18}$O $J=2-1$ peak intensity when there is essentially no gas depletion (i.e., $\delta_\text{gas.ring}=1.0$). This is in contrast to previous studies, which have incorporated high levels of gas depletion in this region ($\delta_\text{gas.ring}=0.025$, \citet{fedele_2017, carney_2018}). This discrepancy may be due to the fact that those models were fit to much lower resolution C$^{18}$O data (beam size $0.36" \times 0.23" \sim 42$ au) compared to the higher resolution data used here ($0.07" \times 0.06" \sim 8$ au). In the dust/gas gap ($r\sim 32$ to $56$ au), our model is broadly consistent with previous works; we find a best-fitting gas depletion factor of $\delta_\text{gas.gap}=0.02$. The full set of parameters from our fiducial model are listed in Table \ref{table:modelparameters}. The initial chemical abundances for the fiducial model and range of values tested in our model grid are listed in Table \ref{table:initial_abundances}. \looseness=-1

\subsection{CS, $^{13}$CS, SO, and H$_2$CS modelling}
\label{subsec:sulfur_results}

Next, we use the CS, $^{13}$CS, SO, and H$_2$CS data to constrain the total gas-phase elemental sulfur abundance S/H. We opt to treat the SiS separately in the next section, since the emission is thought to be tracing shocks in the vicinity of a protoplanet and therefore significantly enhanced from the 'background' level \citep{law_2023}. Using the gas and dust density structures determined in the previous section, we ran models covering a wide range of initial sulfur abundances (S/H $=10^{-12}$ to $10^{-5}$) and C/O ratios (0.3 to 1.1). In each case, the total elemental carbon abundance was fixed at the fiducial value C/H = $1\times 10^{-4}$.

\begin{figure*}
\centering
\includegraphics[clip=,width=1.0\linewidth]{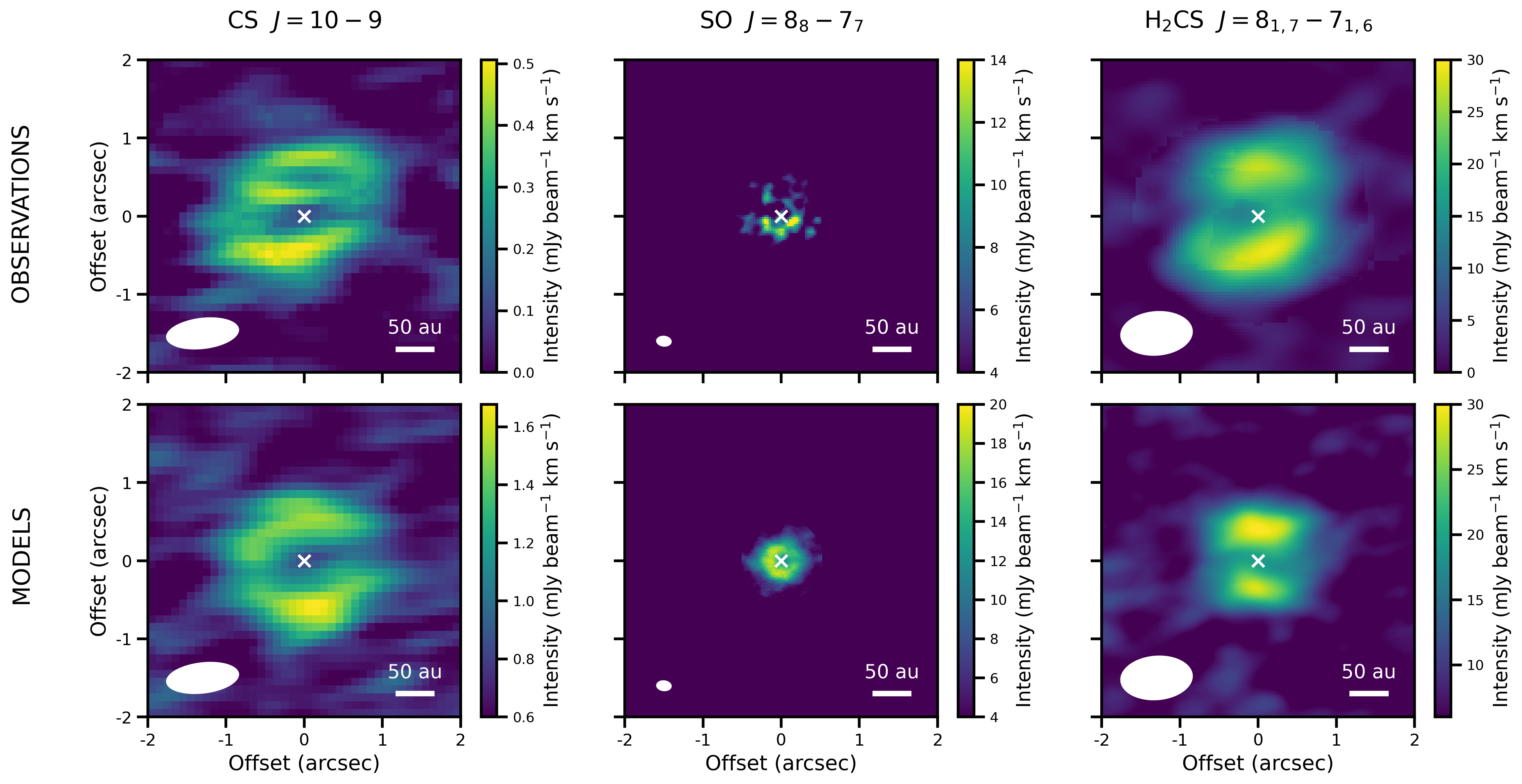}
\caption{Moment 0 integrated intensity maps for CS $J=10-9$, SO $8_8-7_7$, and H$_2$CS $8_{1,7}-7_{1,6}$. \emph{Top row: }ALMA observations. \emph{Bottom row: }Modelled images, creating by processing the \textsc{dali} modelled line image cubes using the CASA tasks \texttt{simobserve} and \texttt{simanalyze}, using parameters matching the observations.}
\label{fig_simalma}
\end{figure*}

Figure \ref{fig_sulfur_grid} illustrates a subset of the results. The top row shows the modelled CS $J=10-9$, $^{13}$CS $J=6-5$, SO $J=8_8-7_7$, and H$_2$CS $J=8_{1,7}-7_{1,6}$ radial intensity profiles, where the initial sulfur abundance is varied and the C/O ratio is fixed at C/O=0.9. The results indicate that total elemental sulfur is heavily depleted from the gas phase by $\sim 3$ orders of magnitude compared to the cosmic abundance, with a best-fit modelled abundance of S/H\;$\sim4 \times 10^{-8}$. This is consistent with observations and modelling of sulfur-bearing species in other disks, which we discuss in detail in Section \ref{sec: discussion}. \looseness=-1

While these models allow us to broadly constrain the total gas-phase sulfur abundance, they provide a poor match to the morphology of the individual radial intensity profiles. The CS $J=10-9$ and H$_2$CS $J=8_{1,7}-7_{1,6}$ observations both peak in the outer dust ring, while our models peak at the central star and fall off gradually to the outer disk. The model provides a slightly better match to the $^{13}$CS $J=6-5$ observation, which is centrally peaked. The models match the SO $J=8_8-7_7$ peak coincident with the inner dust ring, but fail to match the second centralised peak present in the observation. There is also a localised 'bump' in the observed SO $J=8_8-7_7$ within the outer dust/gas gap, which our model again fails to reproduce. Previous studies suggest this component of the emission may be related to ongoing planet formation in the disk \citep{law_2023}, and it is therefore reasonable to expect that our model would not include a feature in this region. 

The second row of Figure \ref{fig_sulfur_grid} shows another subset of models, this time illustrating the impact of varying the C/O ratio while keeping the total elemental sulfur abundance fixed at S/H\;$=4\times 10^{-8}$ and total elemental carbon abundance fixed at C/H = $10^{-4}$. Here, the effect of increasing the C/O ratio leads to an increase in the intensity of the CS, $^{13}$CS, and H$_2$CS lines, but the general morphology of the radial intensity profiles remains the same as before. Interestingly, the intensity of the SO line profiles also increase with increasing C/O until C/O exceeds unity, after which they begin to decrease. This behaviour is somewhat unexpected and is further discussed in Section \ref{sec: discussion}. In general, our results demonstrate that the observed morphological difference between the observations and models cannot be attributed to simple variations in the C/O ratio alone.

In an attempt to better reproduce the emission morphology, we constructed a `composite' model, in which the C/O ratio is allowed to vary radially. This is motivated by an increasing number of studies which suggest high levels of carbon/oxygen depletion and enhanced C/O ratios in the outer regions of protoplanetary disks \citep[e.g.,][]{Du_2015, Bergin_2016, miotello_2019, maps_7_bosman2021, maps8_alarcon2021}, coupled with the inward drift of icy material which acts to enhance the the levels of volatile oxygen in the inner regions \citep{banzatti_2020, gasman_2023}. It is reasonable to expect that such processes may operate in HD\;169142 given its ringed structure, which is indicative of significant physical evolution. In our simple prescription, we divide the disk into two radial regions; an inner region in which the C/O ratio is approximately solar (C/O=0.5), and an outer region in which the C/O region is elevated. Variations in C/O are driven solely by variations in the oxygen abundance, where the total elemental carbon abundance is fixed at C/H $=10^{-4}$. By mimicking the radial depletion and enhancement of volatile oxygen in this way, we are able to explore simple radial variations in the C/O ratio while maintaining a good fit to the CO, [CI], and [OI] data presented in the previous section. 

Results for the composite model are illustrated in the bottom panel of Figure \ref{fig_sulfur_grid}, in which the division point between the two regions is set to the inner edge of the outer dust ring ($r=56$ au). We find that, when the C/O ratio is elevated in the outer disk (C/O $\gtrsim 1$), the model is able to reproduce the peak in the CS $J=10-9$ radial intensity profile, while maintaining the fit to the SO $J=8_8-7_7$ data. The composite model also better matches the H$_2$CS $J=8_{1,7}-7_{1,6}$ observation, with the peak now shifted from the center to the outer dust ring. While this model provides a much better fit overall, the absolute intensities of the CS and H$_2$CS data are still overpredicted when the SO observation is matched, and the modelled $^{13}$CS $J=6-5$ radial intensity profile now incorrectly peaks in outer disk rather than centrally. 

\begin{figure*}
\centering
\includegraphics[clip=,width=1.0\linewidth]{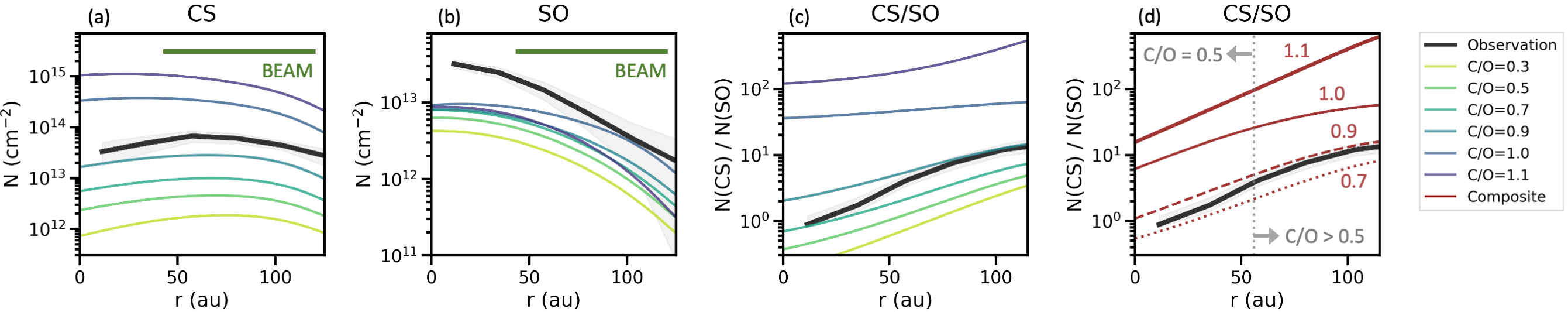}
\caption{CS and SO column densities, and CS/SO column density ratios. Observations in black, with uncertainties shaded in grey. Models in color. All column densities have been convolved with a beam of size matching the CS observation (green bar). \emph{Panel (a)}: CS column densities for models with varying C/O ratio. \emph{Panel (b)}: SO column densities for models with varying C/O ratio. \emph{Panel (c)}: CS/SO column density ratios for models with varying C/O ratio. \emph{Panel (d)}: CS/SO for our composite models using a radially varying C/O ratio. In all cases, S/H is fixed at $4\times 10^{-8}$ and C/H is fixed at $10^{-4}$.}
\label{fig_csso_ratio}
\end{figure*}

We note that a composite model that uses radial variations in the sulfur abundance, rather than the C/O ratio, would not have the same effect. By varying S/H radially, it is possible to better match the emission morphology of a single species, but at the expense of others. For example, adopting a relatively high sulfur abundance of S/H\;$=3\times 10^{-7}$ in the outer disk, while maintaining a lower abundance in the inner disk, allows us to better match the morphology and absolute flux of the CS J=10-9 radial intensity profile. However, this leads to significant over-prediction of SO emission in the outer disk, while the H$_2$CS emission remains under-predicted. It is not possible to simultaneously balance the emission morphology of the CS, $^{13}$CS, SO, and H$_2$CS profiles in this manner. Only through radially varying the C/O ratio are we able to simultaneously improve the fit to the intensity profiles of all species.

We explored varying the radial location at which the two models are spliced together and find that the general morphology can reproduced so long as the splice point lies inside the dust gap ($r \sim 32$ to $56$ au). The precise location within the gap makes has little impact, since the level of gas depletion in the gap is extremely high and does not significantly contribute to the emission of any of the sulfur-bearing molecules.

As an additional point of comparison, we processed the composite model line image cubes using the CASA tasks \texttt{simobserve} and \texttt{simanalyze} to create simulated ALMA observations, using parameters matching the observations. The processed cubes were collapsed to generate moment-zero integrated intensity maps, which we show alongside the observations in Figure \ref{fig_simalma}. These highlight that the emission from species in the composite model provide a close morphological match to the observations, when taking into account the \emph{uv}-coverage and noise associated with the ALMA data.

\subsection{CS/SO ratio}

The CS/SO ratio has been proposed as a tracer of the overall gas-phase C/O ratio in disks, since CS and SO are both highly sensitive to the elemental gas-phase carbon and oxygen abundance \citep[][]{Semenov_2018, maps_12_legal2021} . \citet{booth_2023b} derived the CS/SO ratio for this disk by calculating the CS and SO radial column densities, assuming a fixed temperature of 60 K, finding CS/SO $=0.9\pm0.2$ at 10 au, increasing to $>10$ in the outer disk at 100 au. While these results should be treated with caution due to the assumption of a fixed temperature, comparison with our models can still yield insights into important trends.

We extracted CS and SO radial column densities from a range of models spanning C/O = 0.3 to 1.1. In each case, the initial gas-phase sulfur abundance was fixed at S/H = $4 \times 10^{-8}$, which was chosen since it provides the best match to the SO emission morphology and peak flux, and a close match to the peak fluxes of the other species (albeit not their morphology). We then convolved both column densities to the same angular resolution, matching the beam size of the CS observation (since it is the larger of the two beams). The results are illustrated in Figure \ref{fig_csso_ratio}(a,b), alongside the column densities inferred from the observations by \citet{booth_2023b}. The CS is best matched by the model using C/O = 0.9 and exhibits a similar radial morphology to the observations. The SO column densities are under-predicted by a factor of a few in all cases, but broadly match the morphology of the observations.

The modelled N(CS)/N(SO) ratios are shown in Figure \ref{fig_csso_ratio}(c). The observed profile is best fit with a model using C/O=0.9, which provides an excellent match beyond $\sim 50$ au, but over-predicts CS/SO by a factor of $\sim2$ in the inner disk. Increasing the modelled C/O above unity dramatically over-predicts CS/SO by at least one order of magnitude throughout the entire disk.

Finally, we show the results the for composite models described in the previous section in Figure \ref{fig_csso_ratio}(d). For each model, the C/O ratio is fixed at C/O = 0.5 inside the outer dust ring ($r=56$ au), and is elevated beyond this region. The observations are again best-fit by a model using C/O = 0.9 in the outer disk, but this time provide a slightly closer match in the inner disk compared to models which use a single C/O ratio throughout. Composite models which use C/O$\gtrsim 1$ in the outer disk over-predict CS/SO by at least one order of magnitude. 

As an aside, we also investigated the influence of sulfur abundance on the CS/SO ratio by comparing our fiducial models to those in which S/H is increased by an order of magnitude (Figure \ref{fig_csso_sh}). This analysis demonstrates that variations in the sulfur abundance have a minimal effect, typically altering the CS/SO ratio by less than a factor of 1.5 for a factor 10 increase in S/H. This suggests that the CS/SO ratio is primarily dependent on the C/O ratio, with variations in the sulfur abundance producing only secondary effects.

Our models therefore robustly constrain the C/O ratio in the outer disk ($r\geq 56$ au) to be close to, but not exceeding, unity. The large beam size ($\sim80$ au) limits our ability to precisely constrain the C/O ratio in the inner disk beyond an upper limit of $<1$. However, composite models incorporating an enhancement of volatile oxygen in the inner disk region provide the closest match to the observed CS/SO ratio profile.

\subsection{SiS modelling}
\label{sec:sis_modelling}

SiS emission traces a highly asymmetric azimuthal arc between two dust rings ($r \sim 32$ to $56$ au), and includes an emission 'hotspot' cospatial with the location of the proposed protoplanet HD\;169142b. \citet{law_2023} propose that this emission is the result of a giant embedded planet which produces sufficiently strong shocks to drive a significant amount of silicates into the gas phase. This view is supported by the kinematics, which reveal the SiS emission to be substantially blue-shifted, consistent with a planet-driven outflow.

\begin{figure}
\centering
\includegraphics[clip=,width=1.0\linewidth]{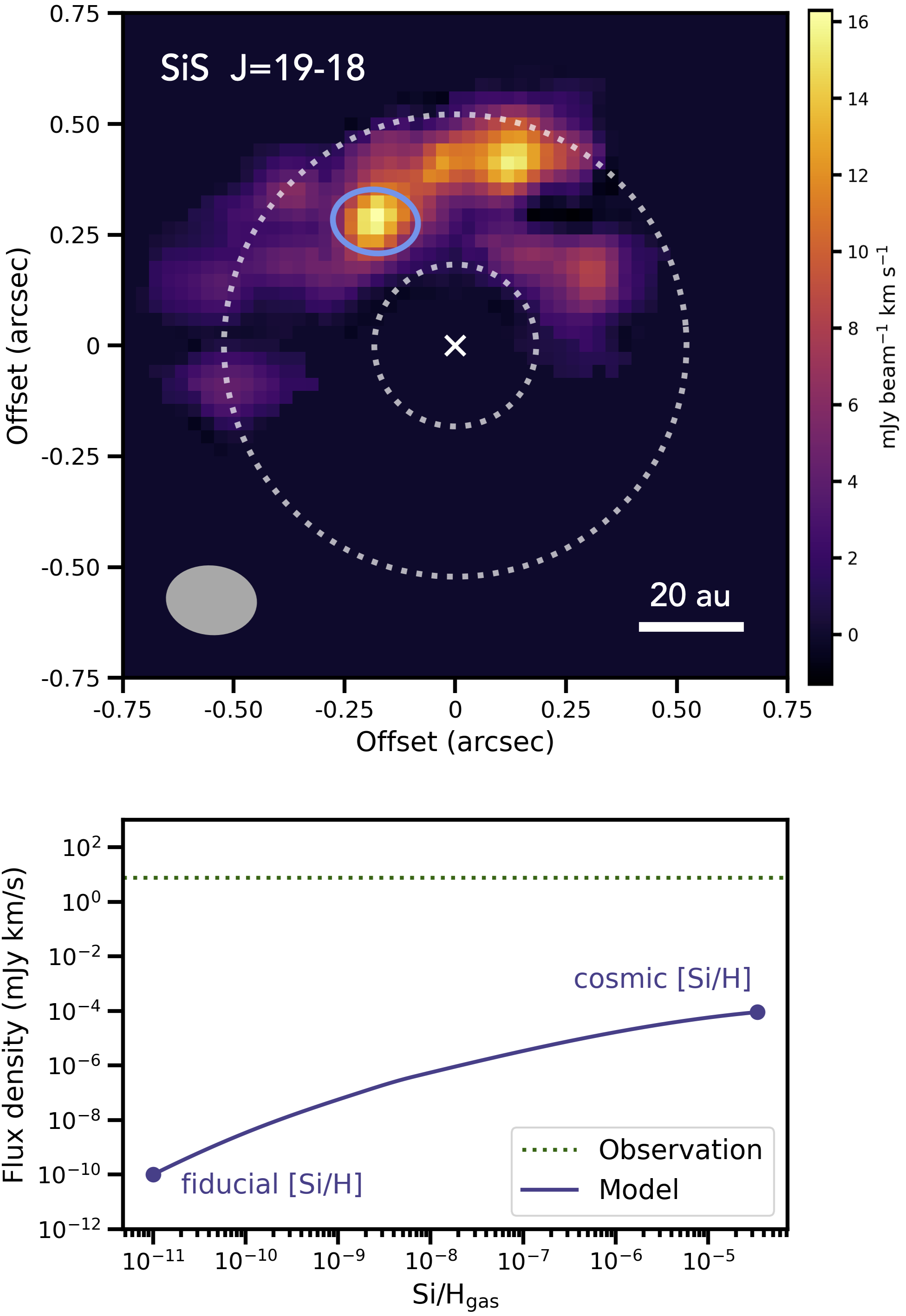}
\caption{\emph{Top: }SiS $J=19-18$ integrated intensity map. Dotted white ellipses denote peaks of the dust rings. The light blue ellipse encompasses the area in which the flux density was measured. The beam size is denoted by the solid grey ellipse in the lower left. \emph{Bottom: } Observed vs. modelled SiS $J=19-18$ flux density, as measured in vicinity of the protoplanet (light blue ellipse in top panel). The models span a range of total elemental Si and S abundances, starting at the fiducial value and increasing up to the cosmic abundance.}
\label{fig_sis_increase}
\end{figure}

In the previous section, we showed that elemental sulfur is highly depleted from the gas phase, finding a best-fit abundance of S/H$=4\times 10^{-8}$. Here, we use our model to determine the SiS $J=19-18$ emission that might be expected under `normal' (i.e., planet-free) disk conditions. We compare this to the observed flux and attempt to quantify the increase in gas-phase SiS due to the protoplanet. For simplicity, we begin by focusing only on the component of emission in the direct vicinity of the planet, measuring the observed flux density contained within one beam (Figure \ref{fig_sis_increase}, top panel). We then extract an SiS $J=19-18$ image cube from the \textsc{dali} model, collapse it to produce an integrated intensity map, and convolve it with the same beam size as the observations. We measure the flux in this region and find a flux density that is many orders of magnitude higher than the model ($\sim7.5$ mJy km/s compared to $\sim 10^{-10}$ mJy km/s). 

Next, we test the effects of increasing the gas-phase silicon and sulfur abundances on the modelled SiS emission. We increase both Si/H and S/H in lockstep, following the ratio of their cosmic abundances [Si/H]/[S/H] = $(3.55\times10^{-5})/(1.86\times10^{-5}) \approx 1.91$ \citep{savage_sembach_1996}. Because silicon is a refractory element, its abundance in disk gas is expected to be strongly depleted with respect to its nominal cosmic value, which (based on solar abundances) is already a factor of 2.5 lower than that of sulfur \citep{asplund_2009}. Since the initial silicon abundance in our model is lower than the initial sulfur abundance, we reach a point at which S/H is equal to the cosmic abundance, while Si/H is still lower than the cosmic abundance. At this stage, we continue to increase Si/H, keeping S/H fixed, until Si/H is also equal to the cosmic abundance. The results are illustrated in Figure \ref{fig_sis_increase} (lower panel). When the gas phase silicon and sulfur abundances both reach their cosmic abundance, the modelled SiS $J=19-18$ emission is still $\sim5$ orders of magnitude lower that the observed value. 

These results highlight the fact that the observed SiS emission is significantly higher than what would be expected under typical conditions in this system. In the model, SiS in the same location as the protoplanet is primarily formed through the following neutral-neutral gas-phase reactions:
\begin{gather}
    \text{Si + SH} \rightarrow \text{SiS + H}
    \label{eq:sis_formation_si_sh} \\
    \text{Si + SO} \rightarrow \text{SiS + O}
    \label{eq:sis_formation_si_so}
\end{gather}
where Equation \ref{eq:sis_formation_si_sh} is the fastest, with a rate coefficient of $\sim 2.5 \times 10^{-10}$ cm$^3$ s$^{-1}$ at the local gas temperature. Our chemical network includes all SiS formation reactions from the UMIST 06 database, supplemented by additional reactions from recent literature \citep{zanchet_2018, mota_2021, galvao_2023}. While it's possible that the network may be missing other important formation reactions, these are unlikely to significantly impact our results, as any additional SiS formed would need to increase the flux density by $\sim 5$ orders of magnitude to match the observations. We further explore this in Section \ref{sec:sis_emitting_area}, where we investigate whether the SiS emission could be related to emission from a circumplanetary disk or shocked gas in the vicinity of a protoplanet.

\section{Discussion}
\label{sec: discussion}

\subsection{Elemental abundance of volatiles}
\subsubsection{Carbon and oxygen}

We have demonstrated that the HD\;169142 disk likely exhibits only moderate depletion of volatile carbon and oxygen compared to their interstellar abundances, with our best-fit model indicating C/H and O/H $\sim 10^{-4}$. We now aim to place this system in context within the broader protoplanetary disk population.

High levels of carbon and oxygen depletion have been inferred across a diverse range of disks. Modelling of the IM Lup disk by \citet{cleeves2018}, for example, showed that observations of C$_2$H are best-fit by a model in which a significant fraction of oxygen is removed from the gas, achieved by depleting H$_2$O by a factor of $\sim 50$ in the warm molecular layer. Later studies by \citet{maps5_zhang2021} and \citet{maps_7_bosman2021} inferred substellar O/H and C/H in the outer disks of AS\;209, HD\;163296, and MWC\;480, with oxygen similarly depleted by factors ranging from 20 to 50. Even higher levels of depletion ($\gtrsim 100$) have been inferred in the outer disks of the T-Tauri systems TW Hya \citep{Kama2016b} and DL Tau \citep{sturm_2022}. This radial dependence is often attributed to the freezeout of key volatiles, which sequester abundant carbon and oxygen carriers in ices at the midplane beyond their respective snowlines. Processes such as turbulent mixing, dust growth, and settling further contribute to locking volatiles from the upper disk atmosphere into midplane ices. Radial drift can then act to transport ices to the inner disk, where they sublimate, enhancing the inner disk volatile reservoir and simultaneously altering the radial C/O ratio \citep{booth_2017}. A potential evolutionary trend has also been proposed, in which the level of elemental gas-phase carbon depletion scales with disk age, supporting the notion that physical and chemical processing of volatiles can sequester carbon and oxygen on relatively short timescales \citep{sturm_2022}.

In contrast, some evolved systems are thought to retain much higher gas-phase carbon and oxygen fractions. Modelling of the well-studied Herbig Ae/Be system HD\;100546, for example, shows that both carbon and oxygen are likely depleted from the gas phase only by a factor of a few \citep{Kama2016b}. Similarly, the disk around DR Tau is thought to be depleted in carbon and oxygen by a factor of $\sim 5$ \citep{sturm_2022}. Our models of the HD~169142 disk align with these results, and interpreting such diversity in the extent of gas-phase elemental depletion between disks across the population is challenging. One possibility is to invoke physical and/or chemical processes which act to slow down or counteract the volatile depletion mechanisms described earlier. A straightforward solution is that some systems are simply too warm to enable large-scale freezeout of CO. Indeed, modelling of HD\;100546 suggests it lacks a significant CO ice reservoir (\citealt{keyte_2023}, \citeblue{Leemker et al. (submitted)}). Alternatively, planetary formation processes may hinder the processing of gas-phase species into ices if a significant proportion of the dust surface area available for freezeout can be removed via transport to the inner disk, accretion, or incorporation into planetesimals. Dust traps associated with newly-forming planets can also produce strong local pressure maxima which inhibit radial drift of ices at the midplane. Additionally, gaps opened by planets can alter the disk thermal structure, locally elevating the gas temperature \citep{chen_2024}. Such processes may be particularly relevant in systems such as HD\;169142 and HD\;100546, both of which have strong observational evidence for multiple planetary candidates \citep[e.g.][]{quanz_2013, quanz_2015, Currie_2015, pinilla_2015, kanagawa_2015, fedele_2017, gratton_2019, hammond_2023}.

An entirely different scenario is proposed by \citet{pascucci_2023}, who contend that large disks are generally not depleted in volatile carbon, CO, or gas. Instead, the high levels of volatile depletion inferred in numerous can be explained models which lack important physics and chemistry, such as the conversion of CO into CO$_2$ ice, and/or sub-millimeter observations which have missing flux due to the absence of short baselines. Future observations with JWST may be able to test this theory, since the shape of mid-IR CO$_2$ ice absorption features can be linked to their formation in water-ice-coated grains \citep{mcclure_2023, sturm_2023}. 

In summary, our models indicate that the HD\;169142 system is only moderately depleted in volatiles, contrasting with higher depletion levels inferred in many other disks. This observed diversity across the population underscores the complexity of the processes governing the chemical composition of such systems, highlighting the need for more complex physical/chemical models and dedicated multi-wavelength observations.

\begin{figure}
\centering
\includegraphics[clip=,width=0.95\linewidth]{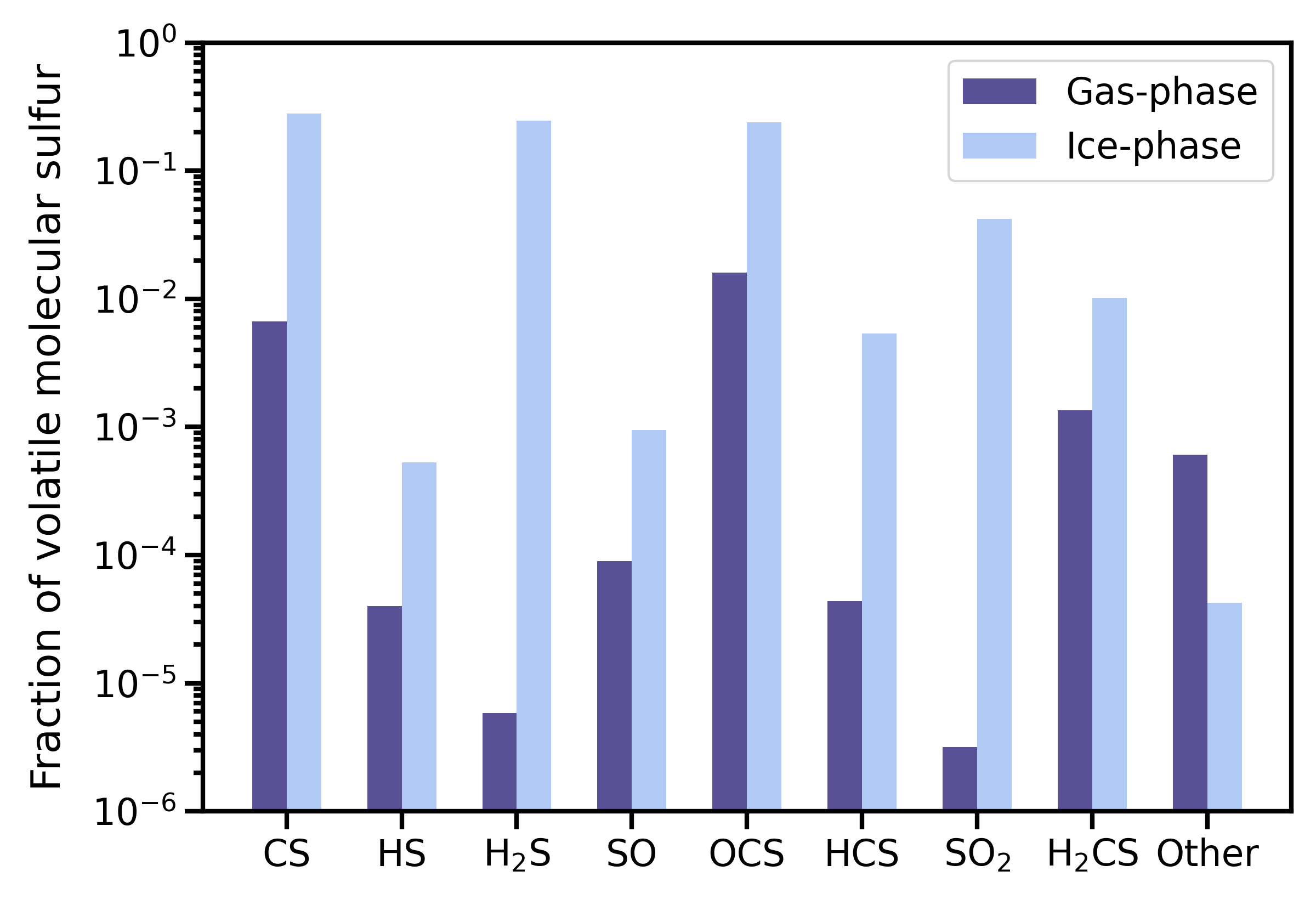}
\caption{Fraction of total molecular volatile sulfur distributed to individual species, summed across the entire disk. Values are extracted from our composite model using C/O=0.5 inside $r=56$ au, C/O=1 outside $r=56$, and a total elemental sulfur abundance S/H=$4\times 10^{-4}$. Dark blue bars correspond to gas-phase fractions and light blue bars correspond to ice-phase fractions.}
\label{fig_fractional_sulfur}
\end{figure}

\subsubsection{Sulfur}

In contrast to carbon and oxygen, elemental sulfur is heavily depleted from the gas phase by a factor of $\sim 1000$. This is consistent with previous studies which suggest that the vast majority of total sulfur in protoplanetary disks is locked into refractory species such as FeS \citep{pasek_2005} or sulfur chains S$_n$ \citep{jimenez_escobar_2011}. \citet{Kama19} measured the refractory sulfur abundance in a sample of disks around Herbig Ae/Be stars to be $89\pm8$\% of total sulfur, suggesting that $11\pm8$\% is available to volatile species. This is supported by studies of gas-phase molecular sulfur-carriers in disks, which have routinely found that it is not possible to provide a full accounting  of the total elemental sulfur budget with the observed molecules. Observations of disks around AB Aur, DM Tau, and GG Tau, for example, found gas-phase CS, SO, and H$_2$S abundances that are factors of 10 to 1000 lower than the total elemental sulfur abundance \citep{fuente_2010, dutrey_2011, phuong_2018}.

In recent years, observations with ALMA have enabled more detailed analyses of the sulfur budget in a small number of disks. \citet{maps_12_legal2021}, for example, used observations of CS and H$_2$CS to infer high levels of gas-phase sulfur depletion in the disk around MWC 480 as part of the MAPS ALMA Large Program ('Molecules with ALMA at Planet-Forming Scales'; \citealt{maps_1_oberg2021}). In that study, the best-fitting model adopted a gas-phase abundance of S/H\;$=8 \times 10^{-8}$, consistent with our findings for HD\;169142. \citet{keyte_2024} showed gas-phase sulfur to be similarly depleted in the HD\;100546 system, with a best-fitting `disk-averaged' abundance of S/H$\sim10^{-8}$. That study also leveraged high resolution SO observations to constrain radial variations in the gas-phase sulfur abundance, showing that S/H varies by $\gtrsim 3$ orders of magnitude across the disk. An analysis of sulfur-bearing molecules in DR Tau by \citeblue{Huang et al. (2024)} found similar variations in the radial sulfur abundance profile and high levels of depletion across the disk. Our results here are therefore consistent with the long-held belief that most elemental sulfur in disks is locked into refractory solids or ices, or incorporated into some as-of-yet unidentified volatile species. 

In Section \ref{subsec:sulfur_results} we highlighted the unusual behaviour of the SO radial intensity profiles in our models, where increasing the C/O ratio leads to an increase in the SO emission. Given that the variations in the C/O ratio are driven by changes in O/H while keeping C/H fixed, we might reasonably expect increasing the C/O ratio (ie. decreasing the total oxygen abundance) to result in a lower SO abundance. To understand why this is not the case, we analysed the the manner in which total elemental sulfur is distributed across the different molecular carriers in each of our models. We find that the fraction of total sulfur incorporated into SO increases with increasing C/O, until C/O $>1$ where it begins to decrease. This is accompanied by a decrease in the fraction of sulfur incorporated into SO$_2$ with increasing C/O. Therefore, we can interpret the rise in the SO abundance as a consequence of limited availability of oxygen for SO$_2$ formation, thus favouring the production of SO. We note that, although the SO abundance increases with increasing C/O, the CS abundance increases at a faster rate, so the CS/SO always increases with increasing C/O. This suggests that while CS/SO reliably tracks the C/O ratio, the additional use of CS/SO$_2$ yields more robust constraints on C/O.

Figure \ref{fig_fractional_sulfur} illustrates the distribution of total elemental sulfur among the major gas- and ice-phase molecular carriers in our composite model. Approximately 99.5\% of molecular volatile sulfur is sequestered in ices, leaving $\sim$0.5\% available for gas-phase chemistry. The observed species CS, SO, and H$_2$CS account for approximately $0.11$\%, while the remaining $0.39$\% is mostly locked into OCS. The H$_2$S gas-phase abundance is significantly lower, contrasting with previous studies which predict it to be the predominant sulfur-carrier in planet forming environments \citep[e.g.][]{pasek_2005}. However, H$_2$S is one of the major ice reservoirs, alongside CS, SO$_2$, and OCS. These findings are consistent with previous modelling of the the Herbig Ae/Be system HD\;100546, where CS, H$_2$CS, and OCS were identified as major gas- and ice-phase sulfur reservoirs \citep{keyte_2024}. More recent observations of HD\;100546 did not detect OCS \citep{booth_2024a}, but upper limits on the OCS abundance derived from the non-detection are within a factor of $\sim3$ of the modelled abundance. As such, OCS cannot therefore be ruled out as an important gas-phase sulfur carrier in that disk, and deeper observations of and OCS and H$_2$S are required to definitively address the sulfur budget in both systems.

\begin{figure}
\centering
\includegraphics[clip=,width=1.0\linewidth]{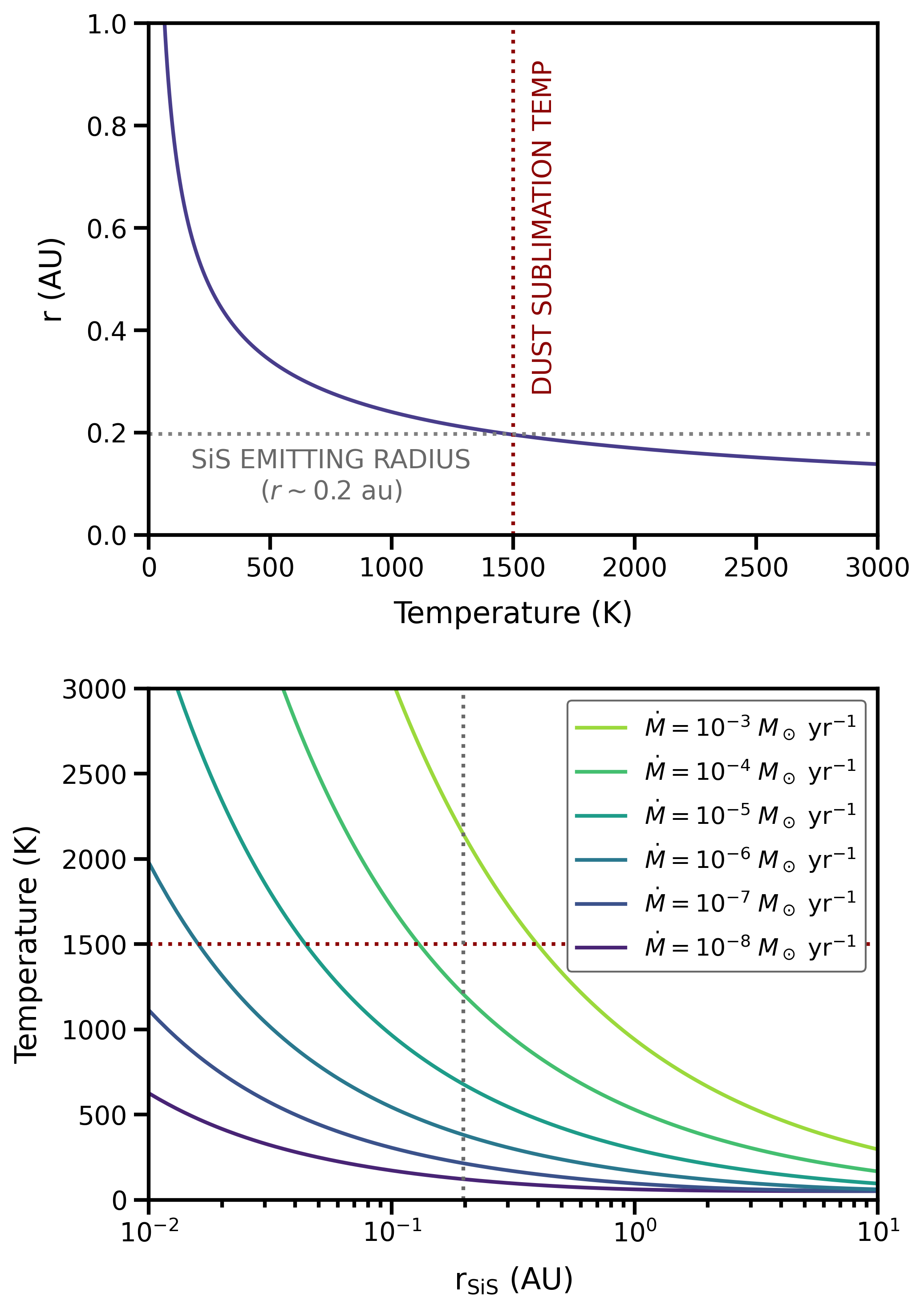}
\caption{\emph{Top panel: }Radius of the SiS $J=19-18$ emitting area as a function of temperature in the optically thick regime. The emitting radius is constrained to $\sim 0.2$ au, using the fact that a grain temperature of ($\sim 1500$ K) would be needed if sublimated silicon was the source of the gas-phase SiS. \emph{Lower panel: } CPD temperature profiles as a function of mass accretion rate. A mass accretion rate of $\dot{M}\sim 10^{-4} M_\odot$ yr$^{-1}$ is required to produce a temperature of $1500$ K at the determined emitting radius of $r\sim 0.2$ au.}
\label{fig_cpd_size}
\end{figure}

\subsection{Linking the SiS emission to a circumplanetary disk}
\label{sec:sis_emitting_area}

In Section \ref{sec:sis_modelling} we showed that our chemical model dramatically underpredicts the observed SiS emission in the vicinity of the proposed protoplanet HD\;169142 b. The modelled SiS emission remains underpredicted by $\sim 5$ orders of magnitude even when the gas-phase atomic sulfur and silicon abundances are artificially increased to their respective cosmic abundances. This suggests that the local gas density and/or temperature must be significantly higher than the 'background' levels predicted by our fiducial model. We now investigate how an embedded planet may impact the disk chemistry, again focusing on the hotspot that is cospatial with the planet. 

Our first aim is to constrain the size of the SiS emitting area. We begin by assuming that the emission is optically thick, such that the intensity of the emission at the line center is given by:
\begin{equation}
    I_\nu = B_\nu (T)
\end{equation}
where $B_\nu (T)$ is the Planck function, and $T$ is the excitation temperature, which is equal to the gas kinetic temperature assuming local thermodynamic equilibrium. We can then relate this to the observed flux density, $F$, by integrating over the linewidth $\Delta \nu$, and solid angle $\Omega$. We measure the flux density in an ellipse matching the beam size centered on the emission peak (Figure \ref{fig_sis_increase}, top panel). Taking the linewidth to be $\Delta \nu = 6$ km/s \citep{law_2023}, the emitting area of the gas as a function of temperature is then given by:
\begin{equation}
    \Omega = \frac{F}{B_\nu (T) \Delta \nu}
\end{equation}
We then convert solid angle to a physical length assuming a source distance of $d=115$ pc \citep{bailer_jones_2021}. Figure \ref{fig_cpd_size} (top panel) shows how the emitting area of the gas varies for temperatures $20-3000$ K. 

Under the assumption that a significant proportion of silicon atoms must be sublimated from dust grain to effectively enhance the gas-phase SiS abundance, we can use the grain sublimation temperature ($\sim 1500$ K) to place an upper limit on emitting area, which we find to be $r \sim 0.2$ au in the optically thick regime.

We now assess whether this result could be attributed solely to emission from a circumplanetary disk. \citet{law_2023} suggest this is unlikely, given that the emission is substantially blue-shifted, possibly tracing a localised outflow. We follow a similar approach to \citet{isella_2014} and \citet{keppler_2019}, where the dust temperature in the CPD at a given radial distance from the planet, $r$, is given by:
\begin{equation}
    T_\text{d}^4(r) = T_{\text{irr}{\star}}^4 (a_\text{p}) + T_\text{irr,p}^4 (r) +T_\text{acc}^4(r)
\end{equation}
where $T_{\text{irr}{\star}}$ is the background temperature of the disk due to heating from the central star (with $a=37$ au, the orbital separation), $T_\text{irr,p}$ is the temperature due to heating from the planet, and $T_\text{acc}$ is the contribution from viscous heating due to accretion in the CPD. 

We adopt a value of $T_{\text{irr}{\star}}=51$ K, which is taken directly from our best-fit model. The irradiation by the planet can be estimated (assuming a CPD aspect ratio of 0.1; \citet{zhu_2018}) as:
\begin{equation}
    T_\text{irr,p}(r) = \bigg(\frac{L_\text{p}}{40\sigma \pi r^2}\bigg)^{1/4}
\end{equation}
where $\sigma$ is the Stefan-Boltzmann constant, and the luminosity of the planet $L_\text{p}$ is taken to be the accretion luminosity:
\begin{equation}
    L_\text{acc} = \frac{G M_\text{p} \dot{M}}{R_\text{p}}
\end{equation}
where $G$ is the gravitational constant, $M_\text{p}$ is the planetary mass, $R_\text{p}$ is the radius of the planet accretion shock front, and $\dot{M}$ is the mass accretion rate of the planet. We set $M_\text{p} = 2.2 M_\text{J}$ based on estimates from observations with SPHERE \citep{gratton_2019}, and $R_\text{p} = 2 R_\text{J}$ which can be considered the typical shock front radius for an accreting planet \citep[e.g.][]{marleau_2022}. Finally, the heating due to accretion is given by:
\begin{equation}
    T_\text{acc}^4(r) = \frac{3 G M_\text{p} \dot{M}}{8\sigma \pi r^3} \bigg[1-\bigg(\frac{r_\text{p}}{r}\bigg)^{1/2}\bigg]
\end{equation}
This treatment is not entirely self-consistent, since we equate the planet's luminosity to its accretion luminosity, and also include an additional accretion-related heating term. However, the contribution to the heating from accretion is relatively minor, and our analysis enables us to place useful upper limits on the CPD temperature when the true luminosity of the planet is unknown.

We compute the temperature profiles for a range of mass accretion rates (Figure \ref{fig_cpd_size}, lower panel). We find that a mass accretion rate of $\gtrsim 10^{-4} M_\odot$ yr$^{-1}$ is required to produce a temperature of $1500$ K at 0.2 au, which is unphysical and many orders of magnitude above what is typically considered a `high' accretion rate for an accreting planet \citep[e.g.][]{lubow_1999}.  This conclusion is not sensitive to the choice of planetary mass, since a realistic accretion rate can only be achieved when $M_\text{p}$ is increased by a factor $\gtrsim 10^3$. Our initial assessment therefore suggests that the SiS emission cannot solely be due to Si sublimation induced by accretion heating in the circumplanetary disk. However, this conclusion rests on our assumption that in order to produce optically thick SiS emission, a significant proportion of Si and S atoms must be liberated from dust via sublimation at temperatures $\gtrsim1500$ K.

\begin{figure*}
\centering
\includegraphics[clip=,width=1.0\linewidth]{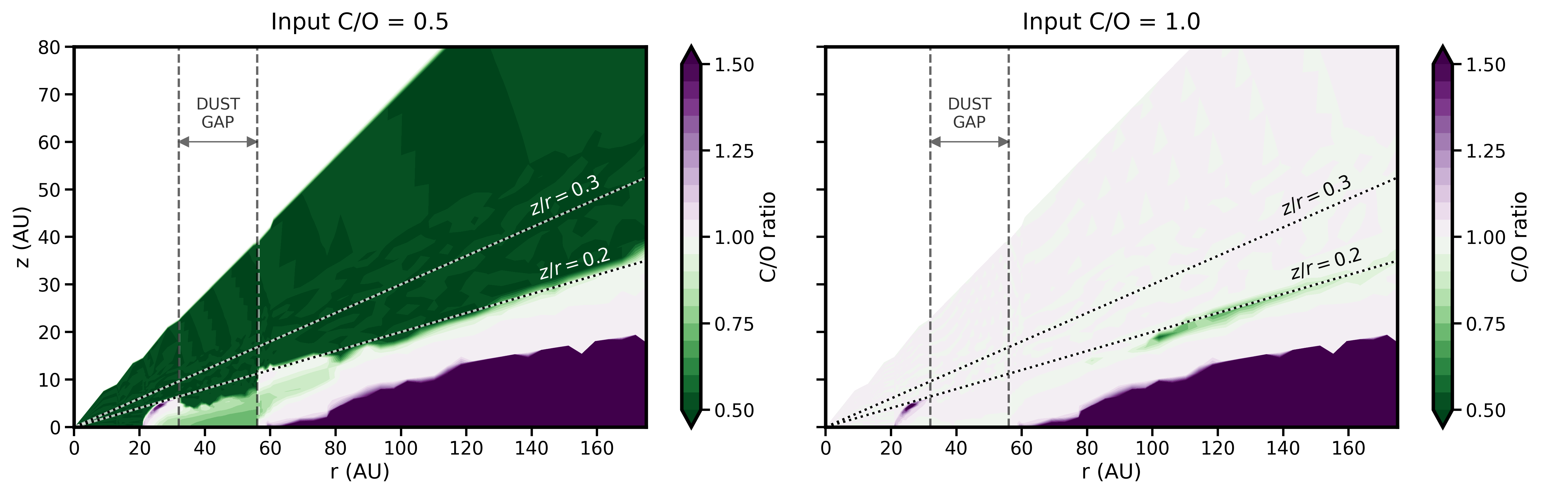}
\caption{Gas-phase C/O ratios for two models. \emph{Left: }Model using input C/O\;=\;0.5. \emph{Right: }Model using input C/O\;=\;1.0. In both cases, freezeout leads to C/O>1 at the midplane beyond $r\sim 60$ au. For the model using input C/O\;=\;0.5, the C/O ratio is also elevated within the dust gap (C/O$\sim$0.75) between the midplane and $z/r \sim 0.1$. Our composite model is constructed using input C/O\;=\;0.5 where $r<56\,$au and C/O\;=\;1.0 where $r\geq56\,$au.}
\label{fig_coratio_map}
\end{figure*}

An alternative scenario is that the SiS emission is related to shocked gas or an outflow driven by an accreting planet. This is supported both by the kinematics and the fact that the emission is not just co-spatial with the planet, but extended around the north side of the disk. Previous research on SiS emission originating from protostellar shocks suggests that SiS is not directly released from grain cores but is likely formed through chemical reactions between neutral molecules at warm temperatures \citep{podio_2017}. If we assume that some sputtering mechanism is able to efficiently liberate silicon atoms from grains back into the gas phase, then SiS may instead be formed through reactions such as Si+SH, which can proceed efficiently in cold gas ($<200$ K) \citep{mota_2021}. We remain agnostic to the exact sputtering conditions. Taking 100 K as a representative value for gas in a CPD accretion shock \citep[e.g.][]{schulik_2020}, we can estimate the emitting area following the same procedure as before, which we find to be $r\sim0.75\,$au based on Figure \ref{fig_cpd_size}. This is comparable to e.g. the massive circumplanetary ring system observed in the transit lightcurve of the star 2MASS\,J140747.7-394542.6, which has an inferred radius of $r_{\rm rings}\approx0.6\,$au \citep{KenworthyMamajek2015}.

Our analysis so far has assumed optically thick line emission. We can approximate the column density required to produce optically thick SiS $J-19-18$ emission by following a similar approach to e.g. \citet{Loomis_2018}. First, we consider the column density of the molecules in the upper state of the transition in the optically thin regime:
\begin{equation}
    N_u^\text{thin} = \frac{4 \pi S_\nu \Delta v}{A_\text{ul}\Omega hc}
\end{equation}
where $S_\nu \Delta v$ is the integrated flux density, $A_\text{ul}$ is the Einstein coefficient (taken from the Cologne Database for Molecular Spectroscopy (CDMS); \citealt{CDMS_muller_2001, muller_2005, endres_2016}), $\Omega$ is the solid angle subtended by the source, $h$ is the Planck constant, and $c$ the speed of light.

Next, we introduce an optical depth correction factor, $C_\tau$, which accounts for the emission not being fully optically thin \citep[e.g.][]{goldsmith_langer_1999}:
\begin{equation}
    C_\tau = \frac{\tau}{1-e^{-\tau}}
\end{equation}
where $\tau$ is the optical depth, which we set to $\tau=1$ to approximate optically thick emission. The optical depth-corrected column density of the molecules in the upper energy state is then given by:
\begin{equation}
    N_u = N_u^\text{thin} \frac{\tau}{1-e^{-\tau}} = N_u^\text{thin} C_\tau
\end{equation}
Finally, the total column density can be calculated:
\begin{equation}
    N_\text{tot} = N_u \frac{Q(T)}{g_u} e^{E_u/T}
\end{equation}
where $Q(T)$ is the molecular partition function (computed by interpolating the values from the CDMS database), $T$ is the gas temperature, and $E_u$ is the upper state energy level (taken from the Leiden Atomic and Molecular Database (LAMDA); \citealt{LAMDA_schoier_2005, klos_lique_2008}).

For an emitting area with radius $r=0.75$ au and gas temperature $T=100$ K (as above), we find $N_\text{tot} = 6.2 \times 10^{16}$ cm$^{-2}$ for optically thick emission. However, if the emission is optically thin, the column density will be significantly lower and the emitting area correspondingly larger. For example, in the optically thin regime, the observed emission intensity can be reproduced using a column density $\sim 5.5 \times 10^{13}$ cm$^{-2}$ if the emitting area is comparable to the size of the synthesised beam (0.19"$\times$0.14"). In the vicinity of he protoplanet, emission outside of one synthesized beam centered on the planet is relatively minor, so this value places a reasonable lower limit on the column density of SiS formed as a result of an accretion shock. This is consistent with the column density of $4.5 \pm 0.7 \times 10^{13}$ cm$^{-2}$ determined by \citet{law_2023}, which was calculated inside a small arc in the north-east of the disk encompassing the emission hotspot in the vicinity of the protoplanet.

\subsection{Composition of planet-accreted gas}
\label{sec:compostion_planet_gas}

Having constrained the gas-phase carbon, oxygen, and sulfur abundances across the disk, we now seek to determine the composition of gas potentially being accreted by the planet. \citep{booth_2023b} previous estimated that the planet is forming within a region where the gas-phase C/O ratio is between $0.8-1.0$, allowing for lower values of C/O if local heating sublimates O-rich ices. Here, we use our chemical model to place tighter constrains on gas-phase C/O in the vicinity of the protoplanet.

Having constrained the gas-phase abundances of carbon, oxygen, and sulfur across the disk, we now aim to determine the composition of the gas that may be accreted by the planet. \citet{booth_2023b} previous estimated that the planet is forming within a region where the gas-phase C/O ratio ranges between 0.8 and 1.0, allowing for lower C/O values if local heating causes the sublimation of oxygen-rich ices. In this section, we utilise our chemical model to establish tighter constraints on the gas-phase C/O ratio in the immediate vicinity of the protoplanet.

We have shown that carbon and oxygen are moderately depleted from the gas phase, and that the gas-phase C/O ratio is generally elevated above solar (C/O\;$\sim1)$. Additional constraints from the sulfur-bearing molecules hint at radial C/O variations, whereby C/O is elevated in the outer disk but approximately solar in the inner disk. However, we are unable to tightly constrain the radial location at which C/O becomes elevated, finding that the data can be well fit when the change occurs anywhere within the dust gap ($r\sim32$ to $56$ au). This is because the main constraints come from the CS emission which peaks in the outer dust ring, and the SO emission which peaks in the inner dust ring. The gas between the two rings is heavily depleted, and therefore does not contribute strongly to emission from either species. Determining the properties of the gas in the vicinity of the protoplanet inside the gap is therefore challenging.

Our aim is to characterise the C/O ratio of gas potentially being accreted by the embedded planet. It is first important to note that our discussion of C/O ratios throughout this study refers to the \emph{input} to our models. In reality, the gas will not be characterised by a single value for C/O, but will vary as a function of radius and height due to freezeout of major volatiles. This is illustrated in Figure \ref{fig_coratio_map}, where we show the gas-phase C/O ratio as a function of radius and height for the two models that are used to construct the `composite' model (with inputs C/O\;=\;0.5 and C/O\;=\;1.0 respectively). We highlight the location of the gap where the protoplanet is located, and the atmospheric layers $z/r=0.2$ and $0.3$, which approximate the location of the warm molecular layer traced by the observations \citep[e.g.][]{yu_2021}.

In both models, C/O at the midplane is above unity beyond $r \sim 60$ au, regardless of the input value. This is driven mainly by freezeout of H$_2$O, where $\gtrsim 75$\% of total elemental oxygen at the midplane is locked into H$_2$O ices. Above $z/r\sim0.2$, C/O is equal to the input value since there is essentially no freezeout in this region. However, within the gap C/O is elevated at the midplane in both models. When the input C/O\;=\;0.5, the gas-phase C/O ratio at the midplane is $\sim0.75$, again mainly due to freezeout of H$_2$O. For the model using input C/O\;=\;1.0, the gas-phase C/O ratio in the gap is also $\sim 1.0$. This suggests that, regardless of which radial location within the gap is chosen as the splice point for our composite model, the C/O ratio of midplane gas in the vicinity of the protoplanet is elevated above solar.

Recent studies suggest that newly-forming planets may infact accrete most of their gas from upper layers of the disk atmosphere via meridional flows \citep[e.g.][]{teague_2019}. If this is the case, then the C/O ratio of the midplane gas may not accurately reflect the composition of the gas being accreted by the planet. Following \citet{cridland_2020}, we can approximate the region of the disk from which gas is accreted to be between one and three scale heights, extending radially to either the planetary Bondi radius or Hill radius, whichever is smallest. The Bondi radius is given by:
\begin{equation}
    R_\text{B} = \frac{GM_\text{p}}{c_\text{s}^2} 
\end{equation}
where $c_\text{s}=\sqrt{k_\text{B} T/ \mu m_\text{H}}$ is the gas sound speed (where $\mu$ is the mean molecular weight and $m_\text{H}$ is the hydrogen mass). Taking $M\text{p}=2.2\;M_\text{J}$, as before, we find $R_\text{B}\approx 2.6$ au. The Hill radius is given by:
\begin{equation}
    R_\text{H} = a \bigg(\frac{M_\text{p}}{3 M_*} \bigg)^{1/3}
\end{equation}
where $a$ is the orbital distance. We find $R_\text{H} \approx 2.8$ au (ie. larger than $R_\text{B}$), and therefore adopt $R_\text{B} = 2.6$ au as the accretion radius.

We calculate the average C/O of the gas between one and three scale heights for both models. For the model with input C/O\;=\;0.5, we find C/O\;$\approx 0.73$ in the planet-accreted region. For the model with input C/O\;=\;1.0, we find that the accreting region also has C/O\;$\approx1.0$. Therefore, despite lack of tight constraints on radial location at which C/O becomes elevated, we can infer that the gas in planet feeding zone has C/O\;$\gtrsim0.73$. 

This analysis provides a snapshot of the current composition of the gas. In reality,  the composition of the gas in the planet's feeding zone will have likely varied across the course of its formation, and will be further complicated if the planet has undergone significant migration. Future studies which couple hydrodynamics and disk chemistry will be able to better constrain the planet's evolutionary accretion history, and test whether the current gas composition accurately reflects the makeup of its atmosphere.

\section{Conclusions}
\label{sec: conclusions}

In this work, we presented new thermochemical models of the HD\;169142 protoplanetary disk. Incorporating a wide range of archival data, we place constraints on the gas-phase elemental carbon, oxygen, and sulfur abundances. We reproduce the spatially resolved emission from a range of sulfur-bearing molecules, and investigate how emission from SiS may be linked to ongoing planet formation in the disk. We also constrain some properties of gas in the planet's accretion zone. Our main conclusions are:

1. The total gas-phase elemental carbon and oxygen abundances are, at most, only moderately depleted from their interstellar abundances. This is in contrast to many systems in which there is evidence for high levels of volatile depletion.

2. Constraints from CO(+isotopologues), [CI], and [OI] observations suggest that the gas phase C/O ratio is elevated above solar, with a best-fit `disk-averaged' value of C/O\;$\sim1$. This is in good agreement with previous studies of this system.

3. Sulfur is highly depleted from the gas phase by a factor of $\sim 1000$. Our best-fitting model predicts S/H\;$=4 \times 10^{-8}$, similar to values inferred in other protoplanetary disks. This provides further evidence for the long-held notion that most elemental sulfur in disks is incorporated into midplane solids, or in volatile form in some unidentified species. Future high-resolution sub-millimeter observations of OCS and H$_2$S will add valuable insights into the volatile sulfur budget of this disk.

4. The emission morphology of sulfur-bearing species CS, SO, and H$_2$CS can only be reproduced when our models incorporate a radial variation in the C/O ratio, where C/O is elevated in the outer disk (C/O > 0.5) but approximately solar in the inner region (C/O$\sim 0.5$). The radial location at which C/O increases is not well constrained ($r\sim 32$ to $56$ au), but approximately coincides with gap between the inner and outer dust rings. This is suggestive of the inward transport of oxygen from the outer disk as midplane ices.

5. The CS/SO$_2$ ratio is a robust tracer of C/O, providing additional constraints when used jointly with CS/SO. Our models show that, while the CS/SO ratio strictly increases with C/O, this does not capture the underlying behaviour of the change in SO abundance. We show that the SO abundance can infact increase with increasing C/O (up to C/O=1), as less oxygen becomes available for SO$_2$ formation and the partitioning of sulfur among the various molecular carriers changes. Conversely, the SO$_2$ abundance constantly decreases with increasing C/O, such that the CS/SO$_2$ ratio varies more dramatically that the CS/SO ratio for changes in C/O.

6. The observed SiS flux is many orders of magnitude above the level predicted from our fiducial model, supporting the theory that is enhanced due to planetary formation processes. We show analytically that the emission is unlikely to result solely from sublimation in a circumplanetary disk, but may be tracing infalling shocked gas.

7. The C/O ratio of the gas in the vicinity of the protoplanet at $r\sim 37$ au is likely elevated above solar (C/O\;$\gtrsim$0.73). Further studies are required to determine whether the current composition of the disk gas reflects the composition of gas accreted over the planets evolutionary lifetime.

\section*{Acknowledgements}

L.K. acknowledges funding via a Science and Technology Facilities Council (STFC) studentship. M.K. gratefully acknowledges funding from the European Union's Horizon Europe research and innovation programme under grant agreement No. 101079231 (EXOHOST), and from UK Research and Innovation (UKRI) under the UK government’s Horizon Europe funding guarantee (grant number 10051045). A.S.B. is supported by a Clay Postdoctoral Fellowship from the Smithsonian Astrophysical Observatory. Support for C.J.L. was provided by NASA through the NASA Hubble Fellowship grant No. HST-HF2-51535.001-A awarded by the Space Telescope Science Institute, which is operated by the Association of Universities for Research in Astronomy, Inc., for NASA, under contract NAS5-26555. M.L. acknowledges support from the Dutch Research Council (NWO) grant 618.000.001. Additionally, M.L. is funded by the European Union (ERC, UNVEIL, 101076613). Views and opinions expressed are however those of the author(s) only and do not necessarily reflect those of the European Union or the European Research Council. Neither the European Union nor the granting authority can be held responsible for them.

\section*{Data Availability}

The data presented in this work are from various programmes detailed in Table \ref{table:observational_data}. The raw ALMA data is publicly available from the ALMA archive.



\bibliographystyle{mnras}
\bibliography{bibliography} 


\clearpage


\appendix

\section{Observational data summary}
A summary of the observational data used in this study is presented in Table \ref{table:observational_data}.

\begin{sidewaystable}[]
\centering
\begin{minipage}{0.85\textheight}
\centering
\caption{Summary of archival observational data used in this study.}
\label{table:observational_data}  
\resizebox{0.85\textheight}{!}{%
\begin{tabular}{l l l l l l l l l}     %
\hline\hline       
                      
Species & Transition & $\nu$ (GHz) \footnote{Line frequencies, Einstein A coefficients (A$_\text{ul}$), and upper energy levels (E$_\text{up}$) are taken from the Cologne Database for Molecular Spectroscopy \citep[CDMS; ][]{CDMS_muller_2001, muller_2005, endres_2016} and the Leiden Atomic and Molecular Database \citep[LAMDA; ][]{LAMDA_schoier_2005}}\label{footnote_cdms} & Beam Size & A$_\text{ul}$ (s$^{-1}$) $^{a}$  & E$_\text{up}$ (K) $^{a}$ & Project \footnote{Observations listed with two project IDs combined both datasets} & PI & Reference \\ 
\hline
\hline
\multicolumn{9}{c}{\emph{ALMA 12m array}} \\
\hline
CONT         & 0.45 mm  & 670.583700 & 0.04"$\times$0.04" (-72$^\circ$)        & -  & - & 2017.1.00727.S & J. Szulágyi & \citet{leemker_2022} \\
CONT         & 0.89 mm  & 331.000000 & 0.05"$\times$0.03" (80$^\circ$)         & -  & - & 2012.1.00799.S & M. Honda & \citet{law_2023}\\
CONT         & 1.3 mm  & 225.000000  & 0.04"$\times$0.02" (73$^\circ$)         & -  & - & 2016.1.00344.S & S. Perez & \citet{garg_2022} \\
CO           & $J=2-1$  & 230.538000 & 0.05"$\times$0.03" (80$^\circ$)         & $6.857\times10^{-7}$   & 16.6 & 2016.1.00344.S & S. Perez & \citet{garg_2022} \\
CO           & $J=3-2$  & 345.795990 & 0.11"$\times$0.09" (74$^\circ$)         & $2.476\times10^{-6}$   & 33.2 & 2012.1.00799.S / 2015.1.00806.S & M. Honda / J.Pineda & \citet{law_2023} \\
$^{13}$CO    & $J=2-1$  & 220.398684 & 0.05"$\times$0.04" (78$^\circ$)         & $6.038\times10^{-7}$   & 15.9  & 2016.1.00344.S & S. Perez & \citet{garg_2022} \\
$^{13}$CO    & $J=3-2$  & 330.587965 & 0.12"$\times$0.09" (75$^\circ$)         & $2.181\times10^{-6}$   & 31.7  & 2012.1.00799.S / 2015.1.00806.S & M. Honda / J.Pineda & \citet{law_2023} \\
$^{13}$CO    & $J=6-5$  & 661.067277 & 0.06"$\times$0.05" (86$^\circ$)         & $1.868\times10^{-5}$   & 111.1  & 2017.1.00727.S & J. Szulágyi & \citet{leemker_2022} \\
C$^{18}$O    & $J=2-1$  & 219.560354 & 0.07"$\times$0.06" (-46$^\circ$)        & $6.011\times10^{-7}$  & 15.8  & 2016.1.00344.S & S.Perez & \citet{garg_2022} \\
{[CI]}     & $^3 P_1 - ^3 P_0$  & 492.160651 & 1.01"$\times$0.51" (81$^\circ$) & $7.880\times10^{-8}$   & 23.6  & 2016.1.00346.S & T. Tsukagoshi & \citet{booth_2023b} \\
CS         & $J=10-9$     & 489.750921 & 0.94"$\times$0.40" (83$^\circ$)       & $2.496\times10^{-3}$   & 129.3  & 2016.1.00346.S & T. Tsukagoshi & \citet{booth_2023b} \\
$^{13}$CS  & $J=6-5$      & 277.455405 & 0.73"$\times$0.46" (89$^\circ$)       & $4.399\times10^{-4}$   & 46.6  & 2018.1.01237.S & E. Macias & \citet{booth_2023b} \\
SO         & $J=8_8-7_7$  & 344.310612 & 0.19"$\times$0.14" (89$^\circ$)       & $5.186\times10^{-4}$  & 87.5  & 2012.1.00799.S / 2015.1.00806.S & M. Honda / J.Pineda & \citet{law_2023} \\
H$_2$CS    & $J=8_{1,7}-7_{1,6}$  & 278.887661 & 0.73"$\times$0.46" (89$^\circ$) & $3.181\times10^{-4}$ & 73.4  & 2018.1.01237.S & E. Macias & \citet{booth_2023b} \\
SiS        & $J=19-18$    & 344.779481 & 0.19"$\times$0.14" (84$^\circ$)       & $6.996\times10^{-4}$   & 165.5  & 2012.1.00799.S & M. Honda & \citet{law_2023} \\

\hline
\multicolumn{9}{c}{\emph{Herschel/PACS}} \\
\hline
HD        & $56 \mu$m    & 5331.56195 & - & $4.860\times10^{-7}$  & 384.6  & DIGIT & N.J. Evans & \citet{kama_2020} \\
HD        & $112 \mu$m   & 2674.98609 & - & $5.440\times10^{-8}$  & 128.5  & DIGIT & N.J. Evans & \citet{kama_2020} \\
{[OI]}    & $63 \mu$m    & 4747.77749 & - & $8.910\times10^{-5}$  & 227.7  & GASPS & W.R.F. Dent & \citet{meeus2012} \\
{[OI]}    & $146 \mu$m   & 2061.06909 & - & $1.750\times10^{-5}$  & 326.6  & GASPS & W.R.F. Dent & \citet{meeus2012} \\
\hline
\end{tabular}
}
\end{minipage}
\end{sidewaystable}


\clearpage

\section{\textsc{dali} model}

\begin{figure*}
\centering
\includegraphics[clip=,width=1.0\linewidth]{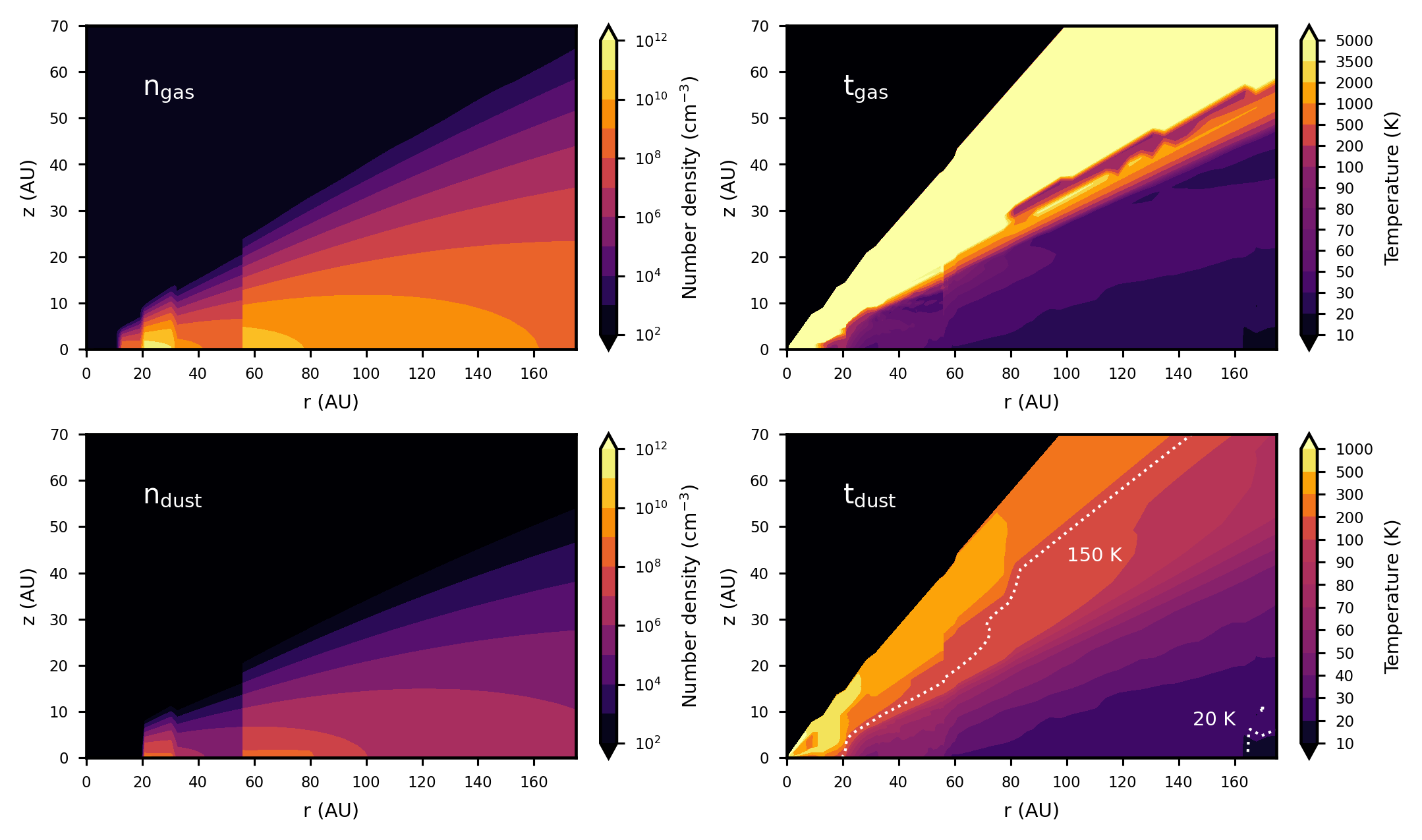}
\caption{Density and temperature structure of the fiducial model. \emph{Top row: }Gas number density (left) and temperature (right). \emph{Bottom row: }Dust number density (left) and temperature (right). The 20 K and 150 K temperature contours are denoted by the dotted white lines, marking the approximate locations of the CO and H$_2$O snowlines.}
\label{fig_tempdens}
\end{figure*}

\begin{figure*}
\centering
\includegraphics[clip=,width=1.0\linewidth]{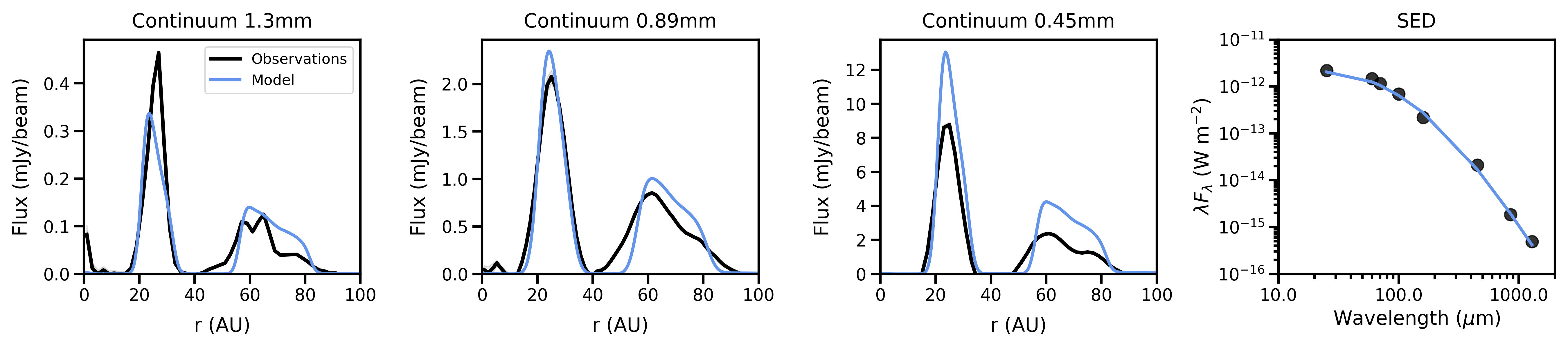}
\caption{Continuum radial intensity profiles and spectral energy distribution (SED). Observations are shown in black and fiducial model in blue.}
\label{fig_continuum_sed}
\end{figure*}

\begin{figure*}
\centering
\includegraphics[clip=,width=1.0\linewidth]{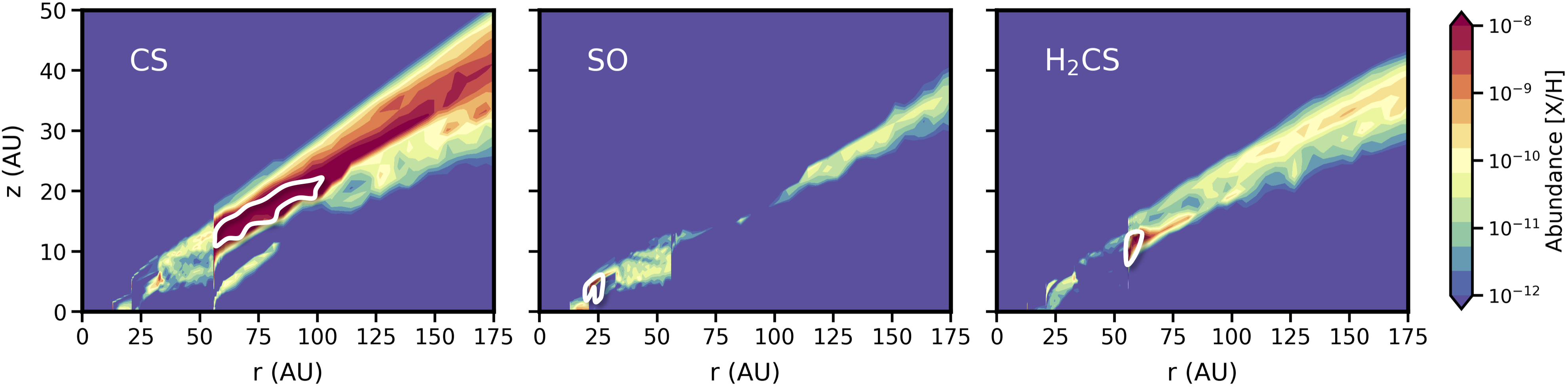}
\caption{CS, SO, and H$_2$CS 2D abundance maps extracted from our composite model (where C/O\;=\;0.5 inside $r<56$ au and C/O\;=\;1.1 outside this region). The white contours denote where 80\% of the emission originates for the CS $J=10-9$, SO $J=8_8-7_7$, and H$_2$CS $J=8_{1,7}-7_{1,6}$ transitions.}
\label{fig_abumaps_sulfur}
\end{figure*}

\begin{itemize}
    \item{Fiducial model parameters list in Table \ref{table:modelparameters}}
    \item Modelled temperature/density maps are presented in Figure \ref{fig_tempdens}.
    \item Modelled continuum radial intensity profiles and SED are displayed alongside the observations in Figure \ref{fig_continuum_sed}.
    \item Modelled CS, SO, and H$_2$CS 2D abundance maps are presented in Figure \ref{fig_abumaps_sulfur}.
\end{itemize}

\begin{table*}
\caption{Fiducial HD~169142 disk model parameters.}             
\label{table:modelparameters}      
\centering
\begin{tabular}{l l l}     %
\hline\hline       
                      
Parameter & Description & Fiducial\\ 
\hline                    
   \rsub                & Sublimation radius                         & 0.2 au    \\
   \rgap                & Inner disk size                            & 0.4 au     \\
   $R_\text{gas.cav}$   & Gas cavity inner radius                    & 13 au      \\
   \rcav                & Dust cavity radius                         & 21 au      \\
   $R_\text{dust.gap.in}$  & Dust gap inner radius                   & 32 au      \\
   $R_\text{gas.gap.in}$   & Gas gap inner radius                    & 32 au      \\
   $R_\text{dust.gap.out}$ & Dust gap outer radius                   & 56 au      \\
   $R_\text{gas.gap.out}$  & Gas gap outer radius                    & 56 au      \\
   $R_\text{L.dust.out}$    & Maximum radius of large dust grains     & 83 au      \\
   $R_\text{out}$       & Disk outer radius                          & 180 au    \\
   $R_c$                & Critical radius for surface density        & 100 au      \\
   $\delta_\text{gas}$  & Gas depletion factor inside $R_\text{gas.cav}$        & $10^{-10}$  \\
   $\delta_\text{gas.cav.edge}$  & Gas depletion factor between $R_\text{gas.cav}$ - \rcav         & $10^{-3}$  \\
   $\delta_\text{gas.ring}$  & Gas depletion factor between \rcav - $R_\text{dust.gap.in}$        & $1.0$  \\
   $\delta_\text{L.dust}$  & Large dust grain depletion factor inside \rcav         & $10^{-10}$  \\
   $\delta_\text{S.dust}$  & Small dust grain depletion factor inside \rcav         & $10^{-10}$  \\
   $\delta_\text{L.dust.ring}$  & Large dust grain depletion factor between \rcav - $R_\text{dust.gap.in}$         & 0.27  \\
   $\delta_\text{S.dust.ring}$  & Small dust grain depletion factor between \rcav - $R_\text{dust.gap.in}$         & 0.27  \\
   $\delta_\text{L.dust.gap}$  & Large dust grain depletion factor between $R_\text{dust.gap.in}$ - $R_\text{dust.gap.out}$         & $10^{-10}$  \\
   $\delta_\text{S.dust.gap}$  & Small dust grain depletion factor between $R_\text{dust.gap.in}$ - $R_\text{dust.gap.out}$          & $10^{-2}$  \\
   
   $\gamma$             & Power law index of surface density profile & 1.0        \\
   $\chi$               & Dust settling parameter                    & 0.2        \\
   $f$                  & Large-to-small dust mixing parameter       & 0.85       \\
   $\Sigma_c$           & $\Sigma_\text{gas}$ at $R_c$               & 8.125 g cm$^{-2}$  \\
   $h_c$                & Scale height at $R_c$                      & 0.07 rad      \\
   $\psi$               & Power law index of scale height            & 0.0       \\
   \gasdust             & Gas-to-dust mass ratio                     & 100        \\
   $L_*$                & Stellar luminosity                         & $10\; L_\odot$        \\
   $T_*$                & Stellar temperature                        & $8400$ K        \\
   $L_X$                & Stellar X-ray luminosity                   & $7.94 \times 10^{28} \text{ erg s}^{-1}$    \\
   $T_X$                & X-ray plasma temperature                   & $7.0 \times 10^{7}$ K     \\
   $\zeta_\text{cr}$    & Cosmic ray ionization rate                 & $1.26 \times 10^{-17}$ s$^{-1}$  \\
   $M_\text{gas}$       & Disk gas mass                              & $2.83 \times 10^{-2}$ \msun   \\
   $M_\text{dust}$      & Disk dust mass                             & $1.11 \times 10^{-4}$ \msun   \\
   $t_\text{chem}$      & Timescale for time-dependent chemistry     & \text{5 Myr} \\
\hline                  
\end{tabular}
\end{table*}

\begin{table*}
\caption{Fiducial HD~169142 disk model initial abundances and range of values covered by the model grid.}             
\label{table:initial_abundances}      
\centering
\begin{tabular}{l l l}     %
\hline\hline       
                      
Species & Abundance (X/H) & Range\\ 
\hline
H    &   1.0                      &   -    \\
He   &   $7.59 \times 10^{-2}$    &   -  \\
C    &   $1.00 \times 10^{-4}$    &   $2 \times 10^{-5}$ to $2 \times 10^{-4}$  \\
O    &   $1.00 \times 10^{-4}$    &   $2 \times 10^{-5}$ to $2 \times 10^{-4}$  \\
N    &   $2.14 \times 10^{-5}$    &   -  \\
S    &   $4.00 \times 10^{-8}$    &   $1 \times 10^{-12}$ to $1 \times 10^{-5}$  \\
Mg   &   $1.00 \times 10^{-11}$   &   -  \\
Si   &   $1.00 \times 10^{-11}$   &   -  \\
Fe   &   $1.00 \times 10^{-11}$   &   -  \\
\hline                  
\end{tabular}
\end{table*}

\section{Dependence of S/H on the CS/SO ratio}

\begin{figure}
\centering
\includegraphics[clip=,width=1.0\linewidth]{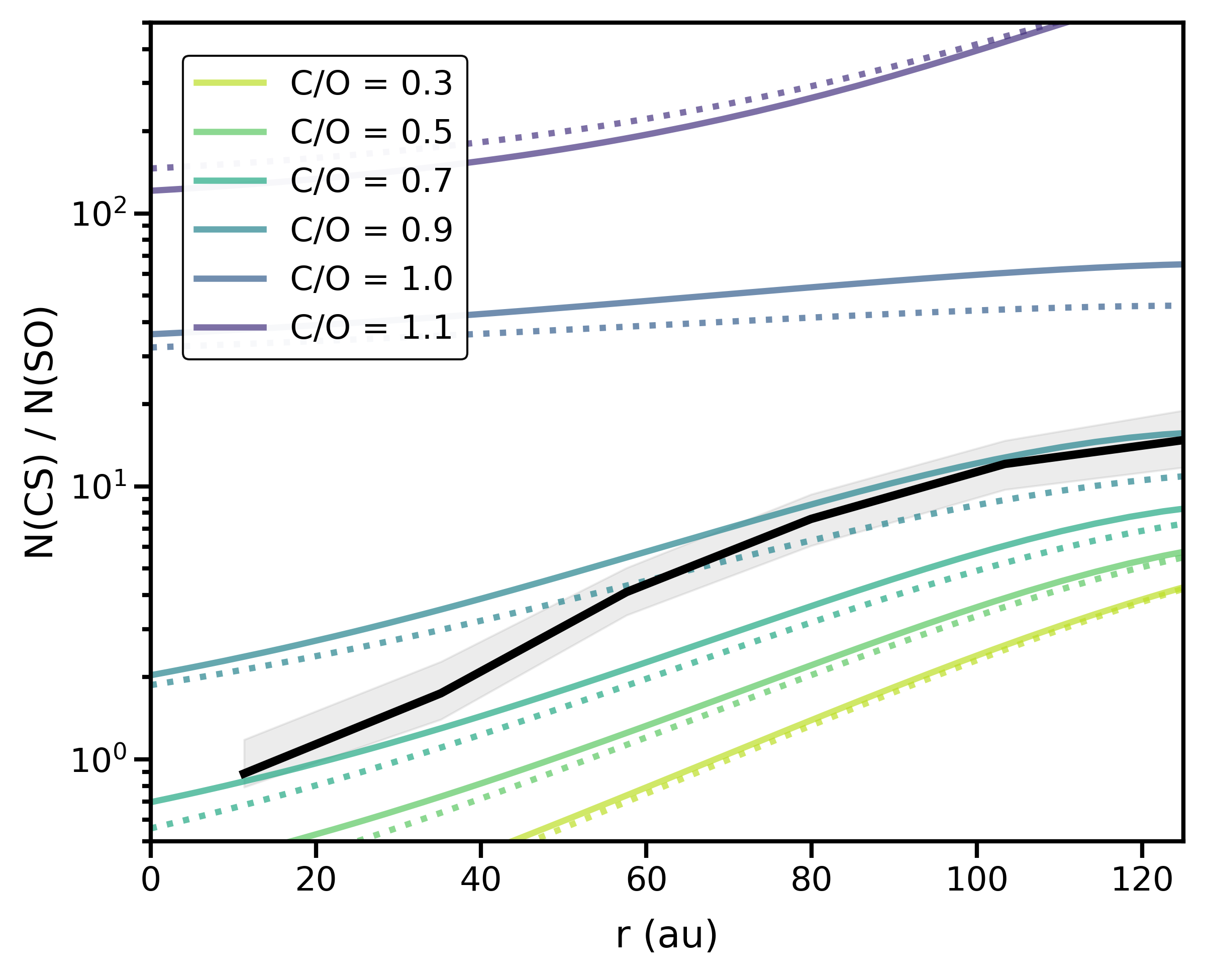}
\caption{Dependence of the CS/SO ratio on the sulfur abundance S/H. Coloured lines indicate models with varying C/O ratio, where solid lines indicate models using S/H$=4 \times 10^{-8}$ and dotted lines indicate models using $4 \times 10^{-7}$. Observed CS/SO ratio shown in black, with uncertainties shaded in grey. The effect of increasing S/H by an order of magnitude is to slightly reduce the CS/SO ratio, typically by a factor $<1.5$.}
\label{fig_csso_sh}
\end{figure}


\bsp	
\label{lastpage}
\end{document}